\documentclass{one_column}
\journal{jas}
\bibpunct{(}{)}{;}{a}{}{,}
\usepackage{csquotes}

\newcommand{\cD}{c_{_\mathrm{D}}}

\renewcommand{\thefootnote}{\fnsymbol{footnote}}
\title{\vspace{5em} Bernhard Haurwitz Memorial Lecture (2017): \\ 
       Potential Vorticity Aspects of Tropical Dynamics}

\author{\Large Wayne~H.~Schubert
\correspondingauthor{\normalsize Wayne~H.~Schubert, Department of Atmospheric Science, Colorado State University, Fort Collins, Colorado 80523}}
\affiliation{Department of Atmospheric Science, Colorado State University, Fort Collins, Colorado
\vspace{30em}\normalfont\normalsize
\begin{displayquote}
This paper is the textual material accompanying the 2017 Bernhard Haurwitz Memorial Lecture, delivered by the author on 28 June 2017, at a joint
session of the American Meteorological Society's 21$^\text{st}$ Conference on Atmospheric and Oceanic Fluid Dynamics and
19$^\text{th}$ Conference on Middle Atmosphere (26--30 June 2017, Portland, OR).
\end{displayquote}
}
\email{waynes@atmos.colostate.edu}

\begin{document}
\maketitle
\clearpage
\setcounter{page}{1}
\renewcommand{\thefootnote}{\arabic{footnote}}

\section{Introduction}                                  %%%%%  Section 1  %%%%%

     Bernhard Haurwitz (1905--1986) was one of the pioneers in atmospheric and
oceanic fluid dynamics, commonly referred to as AOFD. While some of today's
older generation had the good fortune of knowing Bernhard personally, or of
being his colleague, perhaps most AOFD and middle atmosphere researchers know
of his work through the Rossby--Haurwitz wave.
George Platzman \citeyearpar{platzman68} has provided an account of the
somewhat intricate history of the Rossby--Haurwitz wave, during the half
century 1890--1940. An expanded version of Platzman's skeletal history is given
in Table~1.

     As is customary, the history starts with the remarkable work of
Pierre-Simon Laplace (1749--1827).  During the years \citeyearpar{laplace1799}, about a century after Newton's {\it Philosophi\ae\/ Naturalis Principia
Mathematica}, Laplace produced his
monumental, five-volume\footnote{Nathaniel Bowditch published English
translations with extended commentary of the first four volumes from 1829--1839.
Digital versions of these translations are available online at the Internet
Archive.}
treatise entitled {\it Trait\'e de m\'ecanique c\'eleste}, which discusses the
orbits of the then-known planets (out to Uranus), comets, and moons.
In addition, Laplace develops some of the principles of atmospheric and oceanic
fluid dynamics. In Volume I (First Book, Chapter VIII, pages 194--238), Laplace
begins his discussion of the motions of fluids.  On pages 210--212 he considers
the oscillations of a fluid mass surrounding a rotating spheroid. His equation
(325) presents the three components of the vector equation of fluid motion,
including the Coriolis terms\footnote{One wonders if the ``Coriolis terms"
should more accurately be called the ``Laplace terms."} proportional to both
the sine and cosine of the latitude. On page 228 Laplace pauses this discussion
with the statement that ``to determine the oscillations of the sea and the
atmosphere, it is now only necessary to know the forces which act upon these
two fluid masses, and to integrate the preceding differential equations, which
will be done in the course of this work." In Volume II (Fourth Book, Chapters
I--V, pages 526--570), Laplace resumes his discussion of ``The Oscillations
of the Sea and Atmosphere," formulating the required equations when there is
gravitational attraction by the sun and the moon. Although Laplace had a good
understanding of gravity waves, the discovery of planetary waves
is not attributed to him, but rather to Margules (1893) and Hough (1897,
1898), who found that the solutions of Laplace's tidal equations could be
divided into two classes we now call gravitational modes and rotational modes.
Platzman's historical interpretation is as follows:
\begin{displayquote}
``Margules' investigation\footnote{Haurwitz has provided an English translation
of Margules' three part work entitled ``Air Motions in a Rotating Spheroidal
Shell". See the reference list under Margules (\citeyear{margules92},
\citeyear{margules93a}, \citeyear{margules93b}).} of the tidal equations was
the first in which the global planetary wave was explicitly studied from the
standpoint of applications to meteorology.  It was not taken up again from this
point of view until the late 1920's and early 1930's when the Leipzig
school\footnote{A history of the Leipzig school can be found in the book by
Ehrmann and Wendisch \citeyear{ehrmann13}, written on the occasion of the 100th
anniversary of the founding of the Geophysical Institute in 1913.} enlisted it
in an attempt to find a theoretical basis for numerous empirical periodicities
then believed to exist in meteorological data, ranging from a few days to 37
or more days.  The principal theoretical outcome of these efforts was a paper
by Haurwitz in 1937, in which Margules' calculations were extended and improved,
using the more powerful methods developed by Hough. When in 1939 the (Rossby)
trough formula was announced, Haurwitz saw its connection with the theory of
oscillations of the second class that had come down from Margules and Hough,
and through his own hands. He showed how the formula could be extended to allow
for finite width, first on the $\beta$-plane \citep{haurwitz40a} and then on
the sphere \citep{haurwitz40b}. In the latter paper the westward-drift
formula $2\Omega/n(n+1)$ --- first stated by Hough --- was deduced directly for
Rossby's prototype barotropic nondivergent atmosphere, rather than indirectly
from the intricate context of Laplace's tidal equations."
\end{displayquote}

     Since 1940, there have been hundreds of papers written on Rossby--Haurwitz
waves. Table 1 lists the three classics written by Longuet-Higgins
(\citeyear{longuet-higgins64}, \citeyear{longuet-higgins65},
\citeyear{longuet-higgins68}, the latter of which provides a complete solution
of Laplace's tidal equations, including mixed Rossby-gravity waves (also
referred to as Yanai waves) and Kelvin waves.
Thanks to the efforts of many, including \citet{swarztrauber85},
\citet{zagar15}, and \citet{wang16}, the community has software available for
computing the eigenvalues and eigenfunctions of Laplace's tidal equations.

     In the mid-1930's while working on planetary scale problems, Haurwitz also
began thinking about tropical cyclones.\footnote{Some very interesting aspects
of the history and science of tropical cyclones (some before Haurwitz' time)
are discussed in the book by \citet{emanuel05}.}     
Over the years, his interest focused on three topics: 
(i) The motion of binary tropical cyclones \citeyearpar{haurwitz51}; 
(ii) The height of tropical cyclones and of the eye (\citeyear{haurwitz35b}, \citeyear{haurwitz36c}); 
(iii) The Ekman layer under curved air currents 
(\citeyear{haurwitz35a}, \citeyear{haurwitz35c}, \citeyear{haurwitz36b}, \citeyear{haurwitz36a}). 
His work on topic (i) can be considered one of the first efforts at tropical 
cyclone track prediction using simple barotropic arguments.     
His work on topic (ii) was done before there were any 
relevant radiosonde observations or aircraft penetrations and well 
before the era of weather radar and weather satellites. At the time, there 
was uncertainty about the depth of the circulation in tropical cyclones, with 
some arguing that the circulation extended upward only a few kilometers due 
in part to observations of rapid weakening during landfall. Using 
hydrostatic arguments Haurwitz showed that, unless the circulation extends 
up to at least 10 km, the associated temperature distribution is unreasonable. 
Of course, as radiosonde, aircraft, radar, and satellite data became available, 
he was proved correct. Haurwitz also commented on the ``shape of the funnel'', 
which is often referred to as the ``stadium effect'' or the outward slope 
of the eyewall. A modern day example of this is shown in Fig.~1. Haurwitz' work 
on topic (iii) had the goal of generalizing the planetary boundary layer solutions 
of Ekman and Taylor to allow the imposed pressure field to have curved isobars, 
making the solutions more relevant to tropical cyclones.\footnote{For a brief 
summary, see \citet[pages 239--240]{haltiner57}.} 
When asked by \citet{platzman85} to explain what got him interested in this problem, 
Haurwitz replied: ``I wanted to find an explanation why the hurricane has an eye. 
I had in mind that there should be something which comes directly 
out of the hydrodynamic equations, which tells you that the influx 
into the hurricane in the outer part has got to stop somewhere.''  He was on 
the right track, and we shall return to this problem in section 5. However, we
begin our discussion with some more recent concepts about the potential
vorticity structure of tropical cyclones.

\begin{table}[t]                    % Table 1
\caption{A skeletal chronology of the theory of planetary waves. This is a slight 
expansion of the chronology presented by \citet{platzman68}.}
\centering
{\begin{tabular}{cc}  
\hline\hline %\toprule 
   Regional Planetary Waves   &  Global Planetary Waves  \\
   ($\beta$-Plane)            &  (Sphere)  \\
\hline %\midrule
                                                &  \citet{laplace1799} \\
                                                &  \citet{margules92,margules93a,margules93b} \\
                                                &  \citet{hough97,hough98} \\
   \citet{ekman23,ekman32}                      &  \\
   \citet{bjerknes37}                           &  \\
   \citet{rossby39,rossby40}                    &  \\
   \citet{haurwitz37,haurwitz40a}               &  \citet{haurwitz40b} \\
   \citet{longuet-higgins64,longuet-higgins65}  &  \\
   \citet{matsuno66}                            &  \\
   \citet{blandford66}                          &  \\
                                                &  \citet{longuet-higgins68} \\
			                        &  \citet{swarztrauber85} \\
\hline 
%\bottomrule
\end{tabular}}
\vspace*{1em}
\end{table}

\begin{figure}[tb]                                  % Figure 1
\centering
\includegraphics[width=5.5in,clip=true]{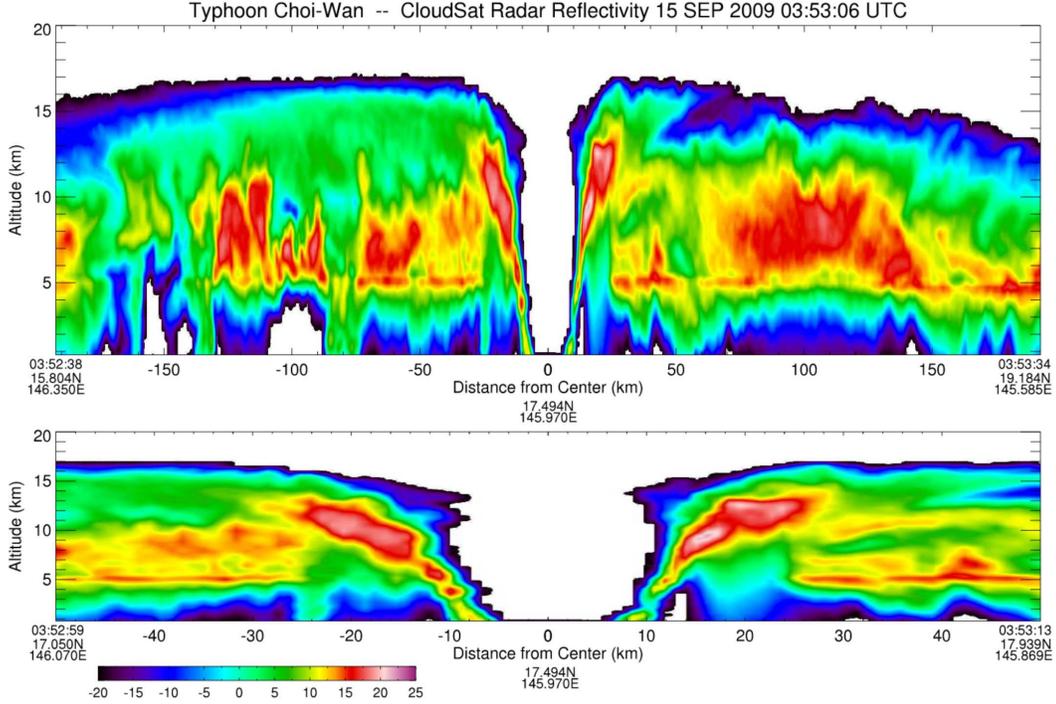}
\caption{At 0353 UTC on 15 September 2009, CloudSat's 94 GHz Cloud Profiling Radar 
passed directly over Typhoon Choi-Wan.  This figure shows a north-south vertical 
cross-section of radar reflectivity for the 65 m s$^{-1}$ storm (north is to the 
right) when it was located approximately 450 km north of Guam.  As is customary, 
in the top panel the horizontal scale is considerably compressed to exaggerate the vertical 
structure.  In the bottom panel, only the region inside a radius of 50 km is shown 
(which excludes the secondary eyewall), but the aspect ratio is one-to-one, thus 
showing the sloping eyewall as would be seen by an observer on a research aircraft. 
This ``stadium effect'' is what Haurwitz referred to as the ``funnel effect.''  
Reflectivity values less than -20 dBZ have been removed for clarity. 
This radar data is courtesy of the NASA CloudSat Project.}
\vspace*{1em}
\end{figure}

\section{The formation of PV towers in tropical cyclones}   %%%%%  Section 2  %%%%%

     It is sometimes said that a tropical cyclone consists of a mesoscale 
($\sim 100$ km) power plant with a synoptic scale ($\sim 1000$ km) supporting 
structure \citep{ooyama82}. Tropical cyclone track forecasting requires that we accurately 
model the synoptic scale supporting structure. The more difficult task of 
tropical cyclone intensity forecasting requires that we accurately model the 
mesoscale power plant. Although there is transient inertia-gravity wave 
activity in a tropical cyclone, the tropical cyclone vortex can be idealized 
to a balanced, PV phenomenon, with the 
balanced wind and mass fields invertible from the PV field. The PV is far 
from being materially conserved. With frictional effects confined primarily  
to a shallow boundary layer, the evolution of PV above the boundary layer 
is determined by advection and diabatic effects. Considering axisymmetric, 
inviscid flow on an $f$-plane, and using radius $r$, height $z$, and time $t$ 
as the independent variables, the PV equation takes the form  
\begin{equation}                                      % Equation (2.1)
       \frac{\partial P}{\partial t}
    + u\frac{\partial P}{\partial r}
    + w\frac{\partial P}{\partial z} 
    = \frac{1}{\rho}\left[
    - \frac{\partial v}{\partial z}\frac{\partial\dot{\theta}}{\partial r}
    + \left(f+\frac{\partial(rv)}{r\partial r}\right)
              \frac{\partial\dot{\theta}}{\partial z}\right],  	  	  
\label{eq2.1}
\end{equation} 
where $u$ is the radial velocity, $v$ the azimuthal velocity, $w$ the vertical 
velocity, $\rho$ the density, $\dot{\theta}$ the diabatic heating, and where 
the potential vorticity is given by 
\begin{equation}                                      % Equation (2.2)
       P = \frac{1}{\rho}\left[
         - \frac{\partial v}{\partial z}\frac{\partial\theta}{\partial r}
         + \left(f+\frac{\partial(rv)}{r\partial r}\right)
                   \frac{\partial\theta}{\partial z}\right]
	 = \frac{fR}{\rho r}\frac{\partial(R,\theta)}{\partial(r,z)},   	  	  
\label{eq2.2}
\end{equation} 
with the potential radius $R$ defined in terms of the absolute angular 
momentum by $\frac{1}{2}fR^2=rv+\frac{1}{2}fr^2$. We will now collapse 
the partial differential equation (\ref{eq2.1}) into two simple ordinary 
differential equations, which can be solved analytically. To accomplish 
this simplification, first transform from the original independent variables 
$(r,z,t)$ to the new independent variables $(R,\theta,\tau)$, where $\tau=t$, 
but $\partial/\partial\tau$ means that $R$ and $\theta$ are fixed, 
while $\partial/\partial t$ means that $r$ and $z$ are fixed. Then, 
the $(r,z,t)$-form of the material derivative is related the  
$(R,\theta,\tau)$-form by 
\begin{equation}                                      % Equation (2.3)
                  \frac{\partial}{\partial t}   
               + u\frac{\partial}{\partial r} 
               + w\frac{\partial}{\partial z} 
               =  \frac{\partial}{\partial\tau} 
         + \dot{R}\frac{\partial}{\partial R}
    + \dot{\theta}\frac{\partial}{\partial\theta},  	  	  
\label{eq2.3}
\end{equation} 
where $\dot{R}$ is the rate that fluid particles are crossing $R$-surfaces. 
For inviscid axisymmetric flow, the absolute angular momentum is materially 
conserved, so that $\dot{R}=0$, which simplifies the right-hand side of 
(\ref{eq2.3}) and hence the left-hand side of (\ref{eq2.1}). The right-hand 
side of (\ref{eq2.1}) can also be simplified by writing it as 
\begin{equation}                                      % Equation (2.4)
      \frac{1}{\rho}\left[
    - \frac{\partial v}{\partial z}\frac{\partial\dot{\theta}}{\partial r}
    + \left(f+\frac{\partial(rv)}{r\partial r}\right)
              \frac{\partial\dot{\theta}}{\partial z}\right]    
     = \frac{fR}{\rho r}\frac{\partial(R,\dot{\theta})}{\partial(r,z)}   
     = \frac{fR}{\rho r}\frac{\partial(R,\theta)}{\partial(r,z)} 
                        \frac{\partial(R,\dot{\theta})}{\partial(R,\theta)}
     = P\frac{\partial\dot{\theta}}{\partial\theta},  	  	  
\label{eq2.4}
\end{equation} 
where the second equality makes use of the Jacobian chain rule and the 
third equality makes use of (\ref{eq2.2}). Then, using (\ref{eq2.3}) and 
(\ref{eq2.4}) in (\ref{eq2.1}), the PV equation becomes 
\begin{equation}                                      % Equation (2.5)
                  \frac{\partial P}{\partial\tau} 
    + \dot{\theta}\frac{\partial P}{\partial\theta} 
    = P\frac{\partial\dot{\theta}}{\partial\theta}.   	  	  
\label{eq2.5}
\end{equation} 
The PV dynamics (\ref{eq2.5}), which is equivalent to (\ref{eq2.1}), can 
now be written in the characteristic form 
\begin{equation}                                      % Equation (2.6)
      \frac{d\ln P}{d\tau} = \frac{\partial\dot{\theta}}{\partial\theta}
                \quad {\rm on} \quad 
      \frac{d\theta}{d\tau} = \dot{\theta},   	  	  
\label{eq2.6}
\end{equation} 
where $(d/d\tau)=(\partial/\partial\tau)+\dot{\theta}(\partial/\partial\theta)$
is the derivative along the characteristic. The reduction of the original 
equation (\ref{eq2.1}) to the pair of ordinary differential equations 
(\ref{eq2.6}) is a significant simplification, especially because these two 
ordinary differential equations can be solved sequentially. For a given 
diabatic heating $\dot{\theta}$, the first equation in (\ref{eq2.6}) gives 
the variation of $P$ along a characteristic, while the second equation gives 
the shape of that characteristic.

     For the example considered here, we assume that $\dot{\theta}$ is 
independent of $\tau$ and has the separable form 
\begin{equation}                                      % Equation (2.7)
   \dot{\theta}(R,\theta) = \dot{\Theta}(R) 
		\sin\left(\frac{\pi(\theta-\theta_B)}{\theta_T-\theta_B}\right),     	  	  
\label{eq2.7}
\end{equation} 
where $\dot{\Theta}(R)$ is a specified function, and where the bottom isentrope 
$\theta_B$ and the top isentrope $\theta_T$ are constants. In the following,   
we choose $\theta_B=300\,$K and $\theta_T=360\,$K, so that the maximum values of 
$\dot{\theta}(R,\theta)$ occur on the $\theta=330\,$K surface. For the mean 
tropical atmosphere, the $\theta=330\,$K isentropic surface is very near the 
425 hPa isobaric surface, which is a reasonable location for the maximum 
in $\dot{\theta}$. However, in the warm-core of an intense tropical cyclone 
the $330\,$K isentropic surface can bend considerably downward, making the assumption 
(\ref{eq2.7}) less tenable. Thus, the example given here should be regarded 
as most applicable to the early phase in the formation of a PV tower.

\begin{figure}[tb]                                  % Figure 2
\centering
\includegraphics[width=5.5in,clip=true]{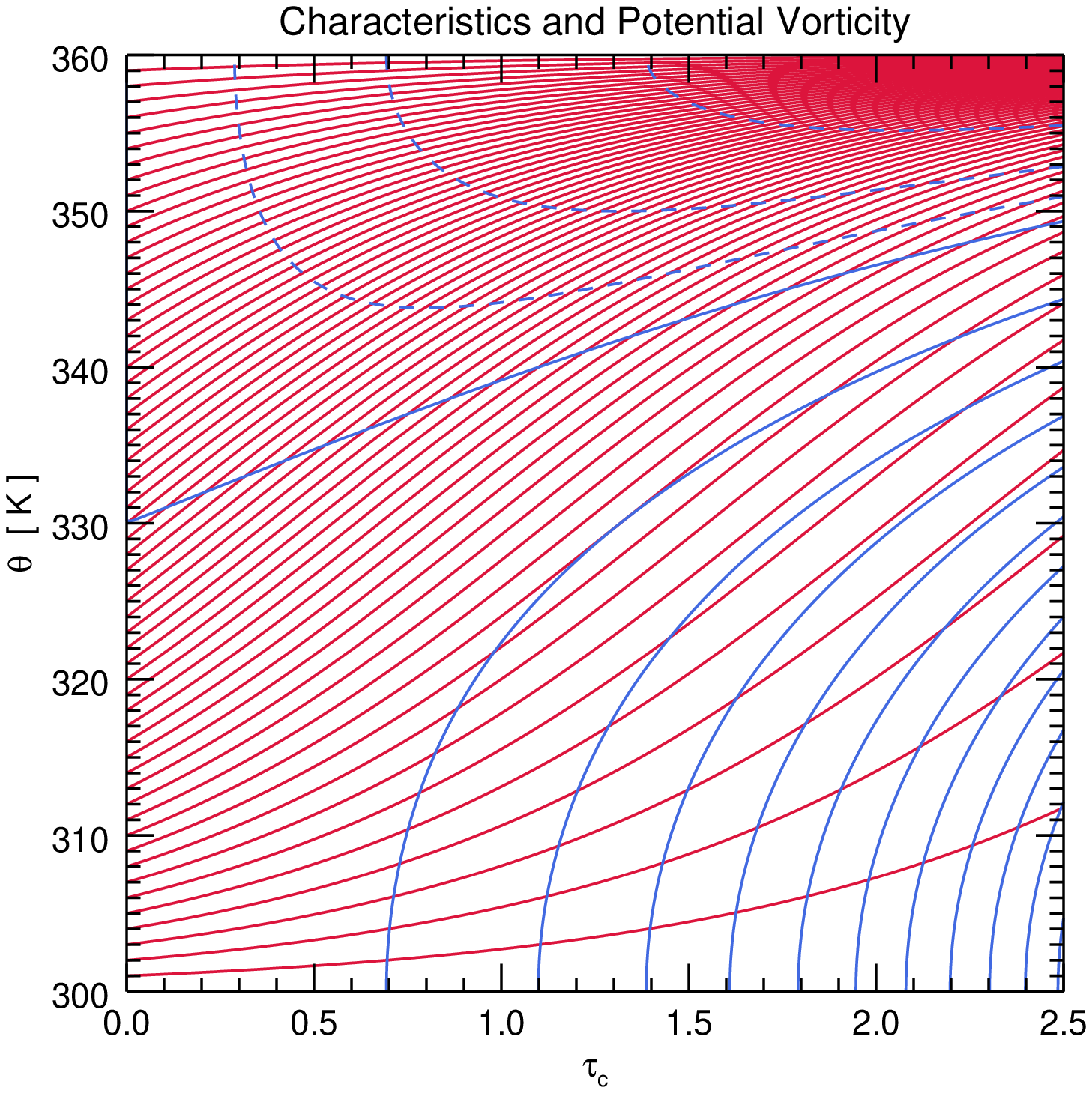}
\caption{The red curves are the characteristics $\theta(\vartheta,\tau_{\rm c})$ 
determined from (\ref{eq2.9}), with $\vartheta$ denoting the characteristic label 
(i.e., the initial value of the potential temperature) and $\tau_{\rm c}$ denoting 
the dimensionless convective clock defined in the second entry of (\ref{eq2.8}). 
The blue curves are isolines of $P(\theta,\tau_{\rm c})/P_0$, as determined from 
(\ref{eq2.12}) and (\ref{eq2.13}). The blue isoline starting at 330\,K corresponds  
to $P(\theta,\tau_{\rm c})/P_0=1$, with the other solid isolines corresponding to 
$P(\theta,\tau_{\rm c})/P_0=2,3,\cdots,11$ and the dashed lines corresponding to 
$P(\theta,\tau_{\rm c})/P_0=1/2,1/4,1/8$.}
\vspace*{1em}
\end{figure}

    Using (\ref{eq2.7}) in the second equation of (\ref{eq2.6}), we obtain 
\begin{equation}                                      % Equation (2.8)
    \frac{\left(\frac{\pi}{\theta_T-\theta_B}\right)d\theta}
         {\sin\left(\frac{\pi(\theta-\theta_B)}{\theta_T-\theta_B}\right)} 
    = d\tau_c,        \qquad \text{where}  \qquad 
        \tau_c = \frac{\pi\dot{\Theta}(R)\tau}{\theta_T-\theta_B}.  
\label{eq2.8}
\end{equation} 
The dimensionless convective clock $\tau_c(R)$ runs at different rates on 
different $R$-surfaces. In 
the eyewall, where $\dot{\Theta}(R)$ is large, the $\tau_c$-clock advances 
quickly, while in the eye and the far-field, where $\dot{\Theta}$ is small, 
the $\tau_c$-clock advances slowly. Note that, for a given $R$, the actual time $\tau=(\theta_T-\theta_B)/\dot{\Theta}(R)$ corresponds to the dimensionless 
convective clock time $\tau_c=\pi$ and can be considered as the tropospheric 
convective overturning time on that absolute angular momentum surface. 
Integration of (\ref{eq2.8}) yields the characteristic equation  
\begin{equation}                                      % Equation (2.9)
   \theta(\vartheta,\tau_{\rm c}) = \theta_B + \frac{2(\theta_T-\theta_B)}{\pi}
      \tan^{-1}\left[e^{\tau_c}  
      \tan\left(\frac{\pi(\vartheta-\theta_B)}{2(\theta_T-\theta_B)}\right)\right], 
\label{eq2.9}
\end{equation} 
where $\vartheta$ is the label of the characteristic, i.e., the initial 
potential temperature of the characteristic. To confirm that (\ref{eq2.9}) 
satisfies the initial condition, note that the $e^{\tau_c}$ factor becomes 
unity when $\tau_c=0$ and that the inverse tangent and tangent operations then 
cancel, so that (\ref{eq2.9}) reduces to $\theta=\vartheta$ when $\tau_c=0$. 
The red curves in Fig.~2 are the characteristics $\theta(\vartheta,\tau_{\rm c})$ 
given by (\ref{eq2.9}), with $\vartheta$ starting at $\vartheta=301\,$K and 
then incremented by $1\,$K up to $\vartheta=359\,$K. Note that these characteristics 
bend upward most rapidly at $\theta=330\,$K, where the value of $\dot{\theta}$ is 
largest. The form (\ref{eq2.9}) is useful for plotting the characteristics because, 
for a given $\vartheta$, it allows for explicit calculation of $\theta$ as 
a function of $\tau_c$. As we shall see shortly, it is also useful to rearrange 
(\ref{eq2.9}) into the form 
\begin{equation}                                      % Equation (2.10)
    \vartheta(\theta,\tau_c) = \theta_B 
        + \frac{2(\theta_T-\theta_B)}{\pi}
	 \tan^{-1}\left[e^{-\tau_c}  
	   \tan\left(\frac{\pi(\theta-\theta_B)}{2(\theta_T-\theta_B)}\right)\right],     	  
\label{eq2.10}
\end{equation} 
which can be regarded as giving the initial potential temperature of the characteristic 
that goes through the point $(\theta,\tau_c)$. 

     We now turn our attention to the solution of the first ordinary differential 
equation in (\ref{eq2.6}). With the previous assumption (\ref{eq2.7}) that 
$\dot{\theta}$ depends only on $(R,\theta)$, and now assuming that the initial 
potential vorticity is the constant $P_0$, the solution of the first equation 
in (\ref{eq2.6}) is 
\begin{equation}                                      % Equation (2.11)
   P(\theta,\tau_c) = P_0 \left(\frac{\dot{\theta}(R,\theta)}
		                     {\dot{\theta}(R,\vartheta(\theta,\tau_c))}\right).    	  	  
\label{eq2.11}
\end{equation} 
One way to confirm that (\ref{eq2.11}) is the required solution is to simply take 
$(d/d\tau)$ of the natural logarithm of (\ref{eq2.11}). Another way is to note that, 
under our assumption that $\dot{\theta}$ is independent of $\tau$, the quantity 
$\dot{\theta}/P$ becomes a Riemann invariant, i.e., $(d/d\tau)(\dot{\theta}/P)=0$, 
from which (\ref{eq2.11}) immediately follows. Using (\ref{eq2.7}) and (\ref{eq2.10}) 
in (\ref{eq2.11}), we obtain the final solution 
\begin{equation}                                      % Equation (2.12)
   P(\theta,\tau_c) = P_0 \left( 
          \frac{\sin\left(\frac{\pi(\theta-\theta_B)}{\theta_T-\theta_B}\right)}
	       {\sin\left\{2\tan^{-1}\left[e^{-\tau_c}  
	           \tan\left(\frac{\pi(\theta-\theta_B)}{2(\theta_T-\theta_B)}\right) 
	       \right]\right\}}\right).    	  	  
\label{eq2.12}
\end{equation} 
Although the right-hand side of (\ref{eq2.13}) is indeterminant at the boundaries 
$\theta=\theta_B,\theta_T$, use of L'Hospital's rule yields 
\begin{equation}                                      % Equation (2.13)
   P(\theta,\tau_c) = P_0 
               \begin{dcases} 
                   e^{-\tau_c}    & \text{if } \theta=\theta_{_T}  \\
	           e^{ \tau_c}    & \text{if } \theta=\theta_{_B}. 
	       \end{dcases}    	  	  
\label{eq2.13}
\end{equation} 
Isolines of the dimensionless potential vorticity $P(\theta,\tau_c)/P_0$ are 
shown by the blue curves in Fig.~2.  The largest value of PV attained on any 
of the 59 characteristic curves (red) occurs at $\tau_{\rm c}=2.5$ 
along the characteristic originating from $\theta=301\,$K. 
This characteristic remains close to the surface where the 
$(\partial\dot{\theta}/\partial\theta)$-term in (\ref{eq2.5}) is large, with 
the result that $P$ acquires the value $11\, P_0$ at $\tau_{\rm c}=2.5$. In 
the region $300 \le \theta\le 330\,$K both the 
$\dot{\theta}(\partial P/\partial\theta)$ and the 
$P(\partial\dot{\theta}/\partial\theta)$ terms in (\ref{eq2.3}) contribute 
to positive $(\partial P/\partial\tau)$. In the region $330 \le \theta\le 360\,$K 
the $P(\partial\dot{\theta}/\partial\theta)$-term makes a negative contribution 
to $(\partial P/\partial\tau)$, but $(\partial P/\partial\tau)$ remains 
positive because of the $\dot{\theta}(\partial P/\partial\theta)$-term. The 
net effect is that a tower of high PV grows into the upper troposphere.

\begin{figure}[tb]                                  % Figure 3
\centering
\includegraphics[width=5.5in,clip=true]{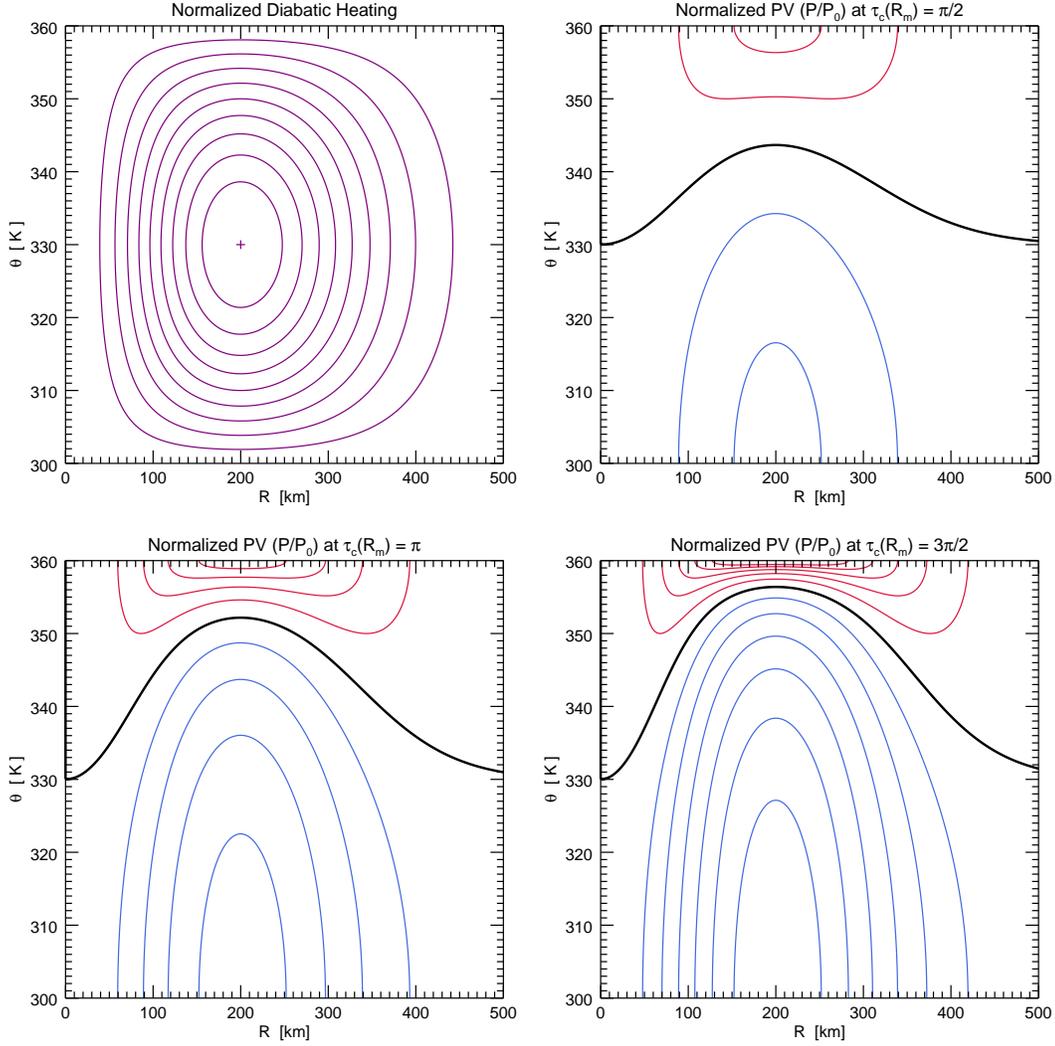}
\caption{The upper left panel shows isolines of 
$\dot{\theta}(R,\theta)/\dot{\Theta}_m$. The remaining three panels show 
isolines of $P(R,\theta,\tau_{\rm c})/P_0$, as determined from 
(\ref{eq2.12})--(\ref{eq2.14}), for $R_m=200$ km and for the three times 
corresponding to $\tau_c(R_m)=\pi/2,\pi,3\pi/2$.}
\vspace*{1em}
\end{figure}

     To get a better appreciation of the formation of a PV tower, and in 
particular the formation of a hollow PV tower, consider the example 
$\dot{\Theta}(R)=\dot{\Theta}_m (R/R_m)^2\exp[1-(R/R_m)^2]$, 
where $R_m$ is the potential radius where $\dot{\Theta}(R)$ is a maximum 
and $\dot{\Theta}_m$ is the value of this maximum. Using this $\dot{\Theta}(R)$ 
in the second entry of (\ref{eq2.8}), we obtain  
\begin{equation}                                      % Equation (2.14)
     \tau_c(R) = \tau_c(R_m) \left(\frac{R}{R_m}\right)^2 
                 \exp\left[1 - \left(\frac{R}{R_m}\right)^2\right], 
              \qquad \text{where} \qquad 
     \tau_c(R_m) = \frac{\pi\dot{\Theta}_m \tau}{\theta_{_T}-\theta_{_B}}.            	  	  
\label{eq2.14}
\end{equation} 
Using (\ref{eq2.14}) in (\ref{eq2.12}), we can now produce $(R,\theta)$-cross-sections 
of PV at different times. Figure 3 shows isolines of PV in $(R,\theta)$-space for the 
choice $R_m=200$ km and for the three times $\tau_c(R_m)=\pi/2,\pi,3\pi/2$. Note 
that a hollow tower of PV is produced.\footnote{For additional discussion, see
\citet{schubert87} and \citet{moller94}.} 

     It is a difficult task to sample a hurricane's wind and mass fields 
in sufficient detail to produce PV maps and PV cross-sections. However, 
using dropsondes and airborne Doppler radar, this has recently been 
accomplished by \citet{bell17} for Hurricane Patricia (2015),  
a storm that rewrote the tropical cyclone record books \citep{rogers17}, 
including the record for the highest surface winds measured by aircraft 
(94 m s$^{-1}$), the highest flight-level winds measured by aircraft 
(99 m s$^{-1}$), and the most rapid intensification rate (54 m s$^{-1}$ in 24 hours). 
One of Bell et al.'s PV cross-sections can be seen in the slide version of 
this paper. It shows a very compact hollow tower of PV, with a maximum value 
of approximately 250 PV units.

\section{Instability of PV rings}                        %%%%%%%%%%  Section 3  %%%%%%%%%%

    Figure 4 is a radar image of Hurricane Dolly (2008) as it approached the south 
Texas coast. The eye is characterized by the low reflectivity blue region and 
the eyewall by the surrounding red region. At this time, the hurricane was estimated 
to have maximum winds of 41 m s$^{-1}$ and a central surface pressure of 976 hPa. 
The wavenumber-4 pattern in the eyewall is believed to be the result of the 
barotropic instability of a thin PV ring. During the past two decades, the instability 
of PV rings has been studied using a hierarchy of models. This hierarchy can be 
conveniently divided into the four levels listed below, starting with the simplest 
nondivergent barotropic models and ending with the most complicated nonhydrostatic, 
full-physics models. 
\begin{figure}[tb]                               % Figure 4 (Hurricane Dolly)
\centering
\includegraphics[width=5.5in,clip=true]{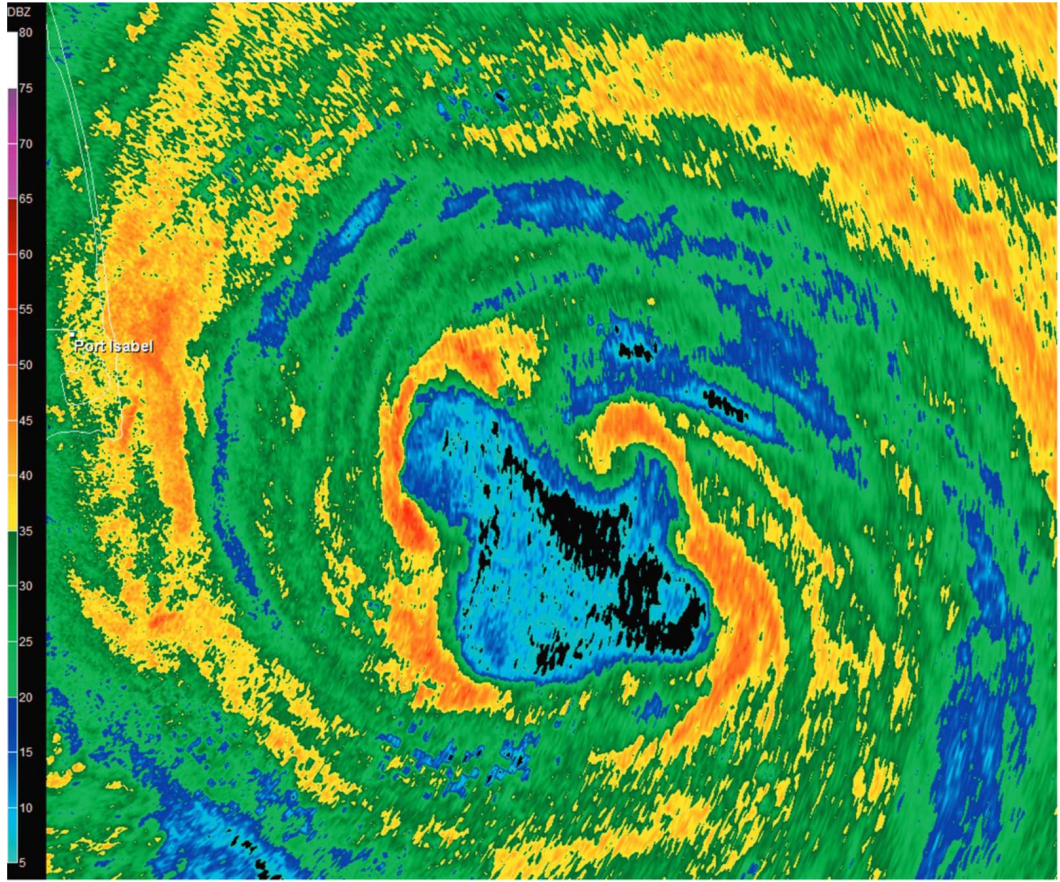}
\caption{Base reflectivity (dBZ color scale on the left) from the Brownsville, 
Texas, National Weather Service radar at 1052 UTC on 23 July 2008. At this time, 
Hurricane Dolly was approaching the coast, and asymmetries, including eye 
mesovortices and straight line segments, were observed in the inner core.
From \citet{hendricks09}.}
\vspace*{1em}
\end{figure}
 
\subsection{Nondivergent barotropic models}
The simplest analysis of PV rings 
involves the eigenvalue-eigenfunction calculation of the exponential instability 
of idealized basic states and then the numerical integration of such small amplitude 
initial unstable flows into their finite amplitude, nonlinear regime 
\citep{schubert99,kossin+schubert01,hendricks09,menelaou13}. 
These studies of the life cycles of unforced, hurricane-like vorticity rings have 
been extended to the forced case by \citet{rozoff09}, who used a 
nondivergent barotropic model to study vortex intensification due to ring forcing. 
The results showed that a smooth intensification can be interrupted in two ways. 
It can be either slowed by a ``vorticity mixing intensification brake'' or enhanced 
by efficient transport of vorticity from the eyewall generation region to the eye.  
Thus, there can be a dual nature of potential vorticity mixing. In addition to 
exponential instability, it is possible for hurricane-like vorticies to exhibit 
algebraic instability, which can occur when the vortex has a maximum in the 
basic-state angular velocity other than at the center of circulation. The dynamics 
of this algebraic instability has been studied in the context of the near-core 
dynamics of hurricanes by \citet{nolan00}.  
 
\subsection{Divergent barotropic models}
The barotropic instability of hurricane-scale 
annular PV rings was first explored by \citet{guinn92} using a nonlinear shallow 
water model that was solved by a normal mode spectral method, thus allowing the total 
flow to be partitioned into its rotational and gravitational components.  	      
As a generalization of the forced nondivergent barotropic model studies discussed 
above, \citet{hendricks14} used a forced shallow water model 
to understand the role of diabatic and frictional effects in the generation, 
maintenance, and breakdown of the hurricane eyewall PV ring. Diabatic heating 
was parameterized as an annular mass sink of variable width and magnitude. The 
mass sink produced a strengthening and thinning PV ring, with dynamic instability 
occurring if the ring became thin enough. The onset of instability was marked by 
drops in both the maximum velocity and the minimum central pressure. 
The most recent research along these lines is that of 
\citet{lahaye16}, who have formulated a unique moist convective shallow water model.

\subsection{Three-dimensional hydrostatic models}
To better understand the vertical 
structure of PV mixing, \citet{hendricks10} used an isentropic 
coordinate model to study the adiabatic rearrangement of hollow PV towers. The model does 
not include diabatic effects, but it shows that in physically relevant cases the 
instability is most vigorous in the lower troposphere.   

\subsection{Three-dimensional nonhydrostatic, full-physics models}
\citet{nolan02} studied the dry dynamics of linearized 
perturbations to hurricane-like vortices. The perturbations were allowed to be 
fully three-dimensional and nonhydrostatic, but for their hurricane-like  
basic states the instabilities were found to be close analogs of those found 
in barotropic models. The appearance of unstable PV rings in nonlinear, 
full-physics models has been documented by several modeling groups, 
including \citet{zhang02}, \citet{yau04}, \citet{kwon05}, \citet{mashiko05},
\citet{wang08,wang09}, \citet{judt10}, \citet{menelaou13b}, 
\citet{naylor14}, and \citet{wu16}. 
The Naylor and Schecter simulations use the CM1 model with  
250 m horizontal resolution, which is more than ten times finer than the 3 km 
resolution often used in WRF simulations. An interesting conclusion of Naylor and Schecter 
is that the initial moist model wave growth leading to polygonal eyewalls and 
mesovortices closely resembles that found in a dry non-convective vortex with the 
same tangential flow. The agreement between the moist and dry models is at least 
partly due to the fact that the bulk of the cloudy eyewall updraft is outside the 
vorticity ring in which the instability occurs. 	      

\vspace*{1em}
     In addition to the above modeling research, there are many observational studies 
that relate to PV rings. These go back to the first reports of polygonal eyewalls by 
\citet{lewis82} and \citet{muramatsu86}. Later, using 
aircraft data, \citet{kossin01} found that the kinematic 
and thermodynamic distributions within the eye and eyewall of strong hurricanes  
evolved between two distinct regimes.  In the first regime, angular velocity was 
greatest within the eyewall and relatively depressed within the eye. In the second 
regime, radial profiles of angular velocity were nearly monotonic with maxima found 
at the eye center. The evolution of these kinematic distributions was often marked 
by a transition from the first regime to the second with the transition possibly 
occurring in less than one hour. Landfalling hurricanes are often well-observed by coastal 
radars and can provide valuable insights, as in the Hurricane Dolly (2008) study of 
\citet{hendricks12} and the Hurricane Ike (2008) study of 
\citet{wingo16}. A case that is under active study is the 
Hurricane Patricia case \citep{rogers17}. Finally, we note that there are 
some very interesting laboratory experiments relevant 
to the instability of PV rings. These have been performed using water as the working fluid \citep{montgomery02} or using a pure electron plasma in a cylindrical trap 
\citep{peurrung93,fine95}.

\section{Subsidence in the eye: hub clouds, eye moats, and warm rings}  %%%%%%%  Section 4  %%%%%%%

\begin{figure}[tb]                               % Figure 5 (Simpson Schematic)
\centering
\includegraphics[width=4.5in,clip=true]{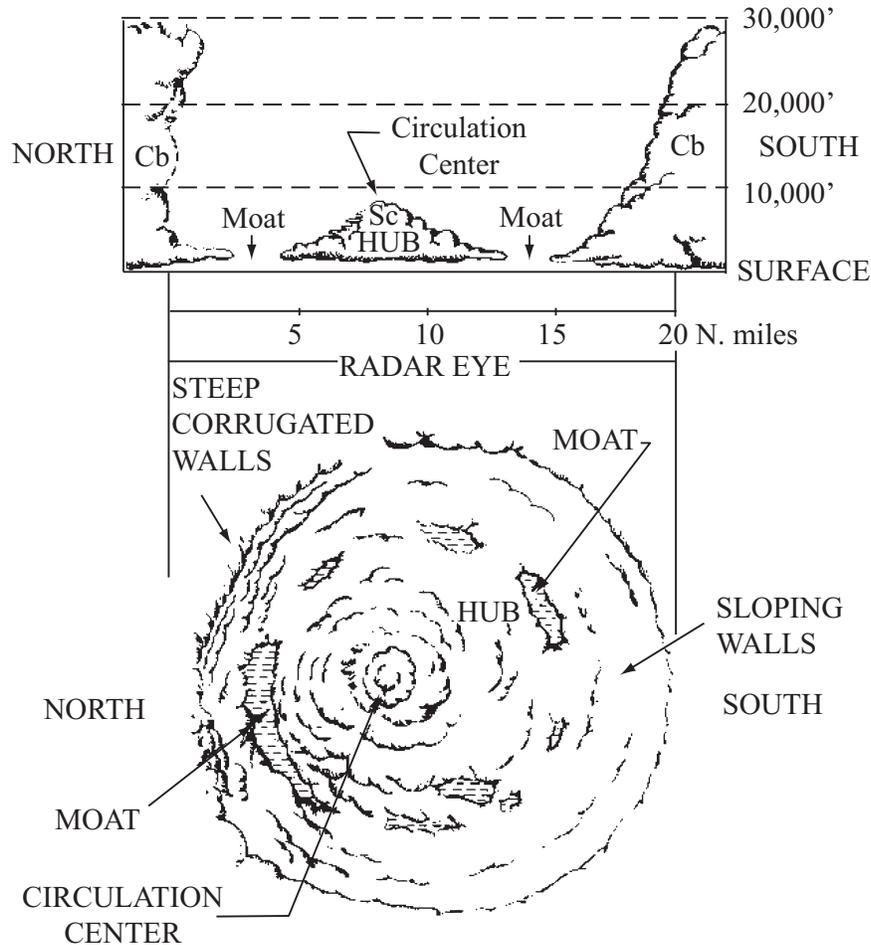}
\caption{Schematic diagram of the eye of Hurricane Edna, 9--10 September 1954, 
as adapted from \citet{simpson55}.}
\vspace*{1em}
\end{figure}
\begin{figure}[tb]                               % Figure 6
\centering
\includegraphics[width=5.0in,clip=true]{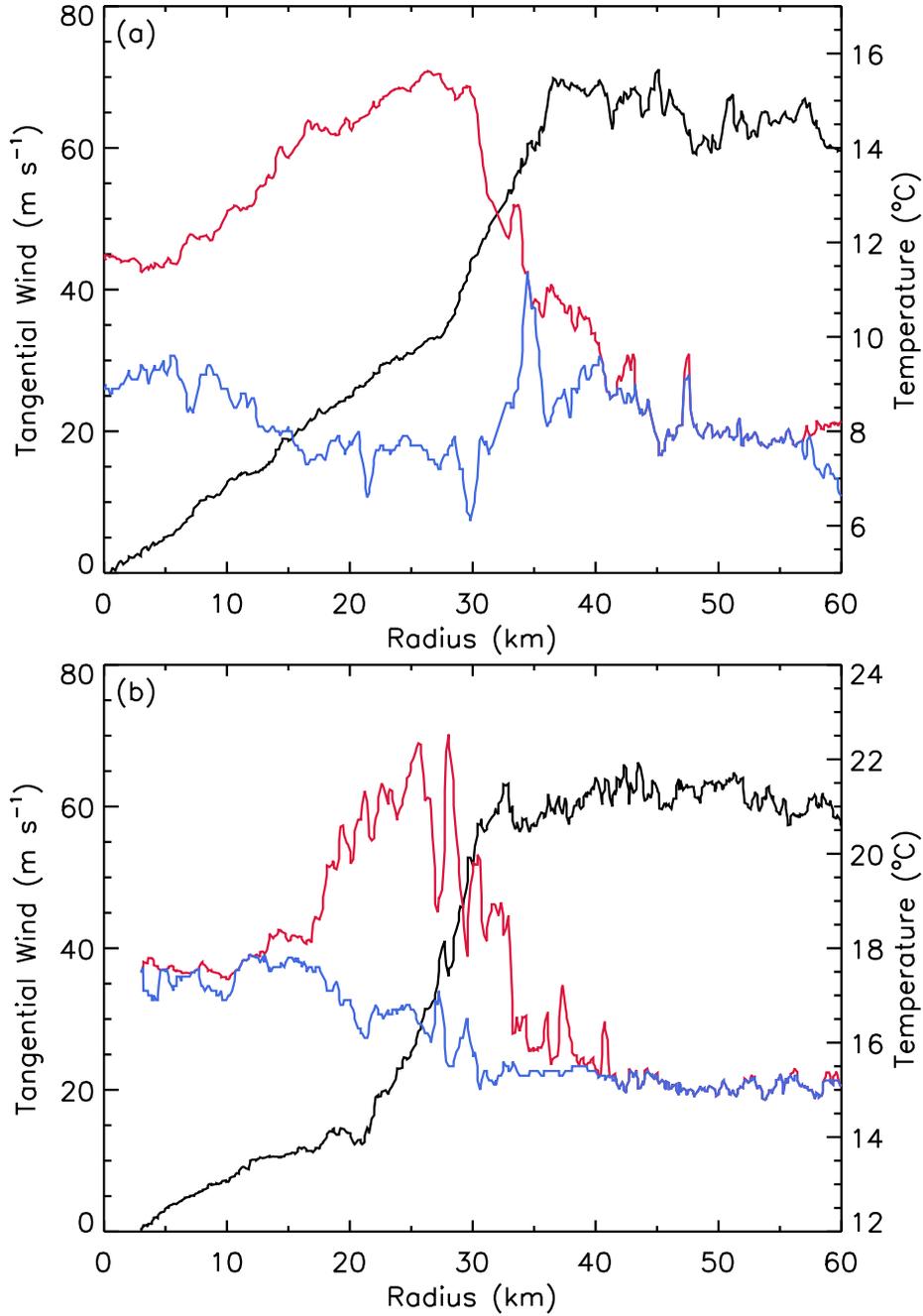}
\caption{Radial profiles of NOAA WP-3D aircraft data for Hurricane Isabel 
on 13 September 2003. The upper panel is for the $z=3.7$ km flight leg 
(1948 to 1956 UTC) and the lower panel for the $z=2.1$ km flight leg 
(1922 to 1931 UTC). Black curves are for tangential wind, red curves 
for temperature, and blue curves for dewpoint temperature. Adapted from \citet{schubert07}.}
\vspace*{1em}
\end{figure}
   The concept of hub clouds and eye moats comes from observations made by 
\citet{simpson55}. Figure 5 is adapted from their 
schematic diagram of Hurricane Edna (9--10 September 1954). Of particular 
interest is the hub cloud near the circulation center and the clear moat 
at the edge of the eye. In later years, intense storms like Edna have been 
found to also possess a warm-ring thermal structure in the lower troposphere. 
A good example is Hurricane Isabel on 13 September 2003, when it had tangential 
winds in excess of 70 m s$^{-1}$.  Figure 6 shows NOAA WP-3D aircraft data for 
this storm.\footnote{For comprehensive discussions of Hurricane Isabel, see 
\citet{montgomery06}, \citet{aberson06}, \citet{schubert07},
\citet{bell08}, and \citet{nolan+stern+zhang09,nolan+zhang+stern09}. 
The eye-moat and mesovortex structure of Hurricane Isabel on the previous 
day (12 September) is discussed by \citet{kossin04} 
and \citet{rozoff06}.} 
The two panels show tangential wind (black lines), temperature (red lines), 
and dewpoint temperature (blue lines) for a 2.1 km altitude radial leg 
(lower panel) and a 3.7 km altitude radial leg (upper panel). A warm-ring 
thermal structure is seen at both altitudes, with the ring approximately 
4.2$^\circ$C warmer than the vortex center at 3.7 km and 5.0$^\circ$C 
warmer than the vortex center at 2.1 km. As can be seen from the dewpoint 
depressions, the warm-ring region near 25 km radius is associated with 
dry, subsiding air at the outer edge of the eye. Enhanced subsidence at 
the outer edge of the eye tends to produce an eye-moat. 
The photograph shown here as Fig.~7 was taken from the WP-3D aircraft near 
the edge of the eye, looking towards the hub cloud at the center of the eye. 
The top of the hub cloud is near 3 km altitude, so the radial leg in the upper   
panel of Fig.~6 is just above the top of the hub cloud, while the radial leg 
in the lower panel is just below the top of the hub cloud, as is evident in 
the dewpoint depressions.

\begin{figure}[tb]                               % Figure 7 (Isabel Photo)
\centering
\includegraphics[width=5.5in,clip=true]{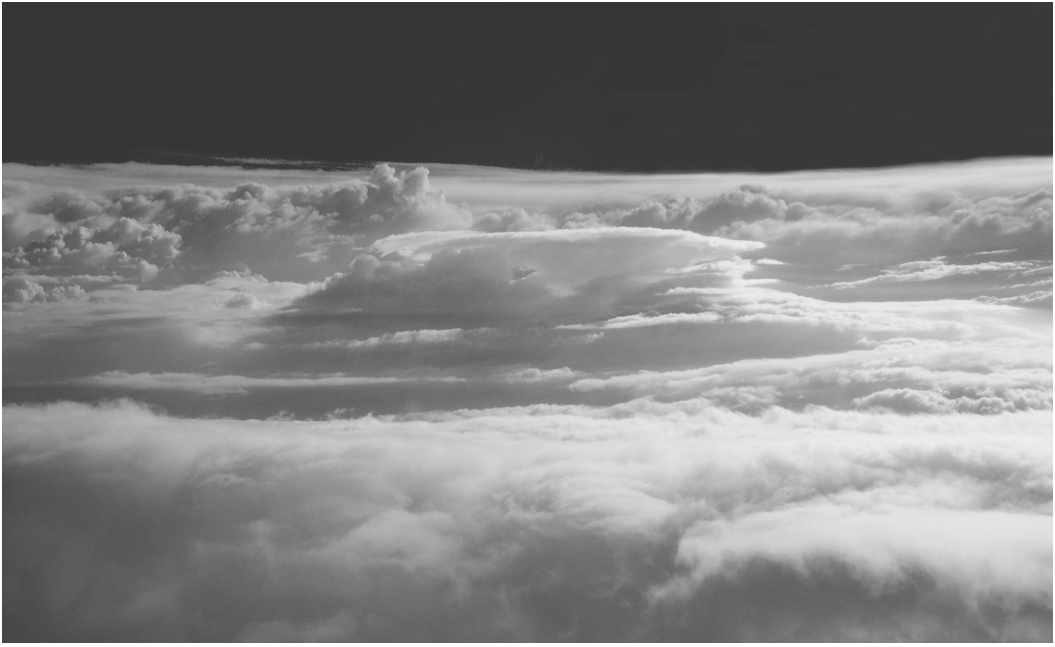}
\caption{Photograph of the eye of Hurricane Isabel on 13 September 2003. 
The 3 km tall hub cloud at the center of the eye is surrounded by a moat 
of clear air or shallow stratocumulus. Beyond the hub cloud and on the 
opposite side of the eye (at a distance of $\sim 60$ km) lies eyewall 
convection extending up to 12--14 km. Since the first internal mode Rossby 
length in the eye of Isabel at this time is approximately 13 km, the eye 
diameter is approximately 5 Rossby lengths, allowing for the rare opportunity 
to view balanced dynamical structure over several Rossby lengths in a single 
photograph. Photo courtesy of Sim Aberson.}
\vspace*{1em}
\end{figure}

     For simplicity, the analysis in this section considers an axisymmetric, 
balanced flow in the inviscid fluid that lies above the frictional boundary 
layer. To simplify the primitive equation model to a balanced vortex model,
we assume that the azimuthal flow remains in a gradient balanced state, 
i.e., we discard the radial equation of motion and replace it with the 
gradient balance condition given below as the first entry in (\ref{eq4.1}).  
A sufficient condition for the validity of this assumption is that the 
diabatic forcing effects have slow enough time scales that significant, 
azimuthal mean inertia-gravity waves are not excited. We shall describe 
this inviscid flow using the log-pressure vertical coordinate 
$z=H\ln(p_0/p)$, where $H=R_d T_0/g$ is the constant scale height,  
$p_0$ and $T_0$ are constant reference values of pressure and temperature, 
$R_d$ is the gas constant for dry air, and $g$ is the acceleration of gravity. 
Under the balance condition, the governing equations are  
\begin{equation}                              % Equation (4.1)
  \begin{split}                        
      & \left(f + \frac{v}{r}\right)v = \frac{\partial\phi}{\partial r}, \qquad        
                  \frac{\partial v}{\partial t} 
               + w\frac{\partial v}{\partial z} 
      + \left(f + \frac{\partial(rv)}{r\,\partial r}\right)u = 0,  \qquad    
      \frac{\partial\phi}{\partial z} = \frac{g}{T_0}T,    \\[2.5ex]        
        &\quad \frac{\partial(e^{-z/H}ru)}{r\,\partial r}
        + \frac{\partial (e^{-z/H} w)}{\partial z} = 0,  \qquad    
               \frac{\partial T}{\partial t} 
           + u \frac{\partial T}{\partial r} 
           + w \left(\frac{\partial T}{\partial z} + \frac{RT}{c_p H}\right) 
	   = \frac{Q}{c_p},  
  \end{split}
\label{eq4.1}
\end{equation}
where $f$ is the constant Coriolis parameter, $\phi$ the geopotential, $u$ the 
radial velocity component, $v$ the azimuthal velocity component, $w$ the 
log-pressure vertical velocity, and $Q$ the diabatic heating. 

     The thermal wind equation is derived by eliminating $\phi$ between the 
hydrostatic equation and the gradient wind equation. Then, taking 
$\partial/\partial t$ of this thermal wind equation, we obtain 
\begin{equation}                                     % Equation (4.2)
    \frac{\partial}{\partial z}
    \left[\left(f + \frac{2v}{r}\right)\frac{\partial v}{\partial t}\right] 
  = \frac{g}{T_0}\frac{\partial}{\partial r}\left(\frac{\partial T}{\partial t}\right),    
\label{eq4.2}
\end{equation}
which is the constraint that must be satisfied by the local time derivatives of the 
azimuthal wind field and the temperature field. Using the mass conservation principle,
we define a streamfunction $\psi$ such that 
\begin{equation}                                    % Equation (4.3)
        e^{-z/H} u = -\frac{\partial  \psi }{   \partial z},  \qquad  
        e^{-z/H} w =  \frac{\partial(r\psi)}{r\,\partial r}.   
\label{eq4.3}
\end{equation}
Using (\ref{eq4.3}) in the second and fifth entries of (\ref{eq4.1}), we obtain 
\begin{equation}                                   % Equation (4.4)
      -\left(f + \frac{2v}{r}\right)\frac{\partial v}{\partial t}
    + B\frac{\partial(r\psi)}{r\,\partial r} 
    + C\frac{\partial\psi}{\partial z} = 0,   \qquad 
        \frac{g}{T_0}\frac{\partial T}{\partial t}
     + A\frac{\partial(r\psi)}{r\,\partial r} + B\frac{\partial\psi}{\partial z} 
     = \frac{g}{c_p T_0}Q,         
\label{eq4.4}
\end{equation}
where the static stability $A$, the baroclinicity $B$, and the inertial 
stability $C$ are given by   
\begin{equation}                                   % Equation (4.5)
  \begin{split}
      e^{-z/H} A&= \frac{g}{T_0}\left(\frac{\partial T}{\partial z}+\frac{R_d T}{c_p H}\right), \qquad 
      e^{-z/H} B =-\left(f + \frac{2v}{r}\right)\frac{\partial v}{\partial z} 
                 =-\frac{g}{T_0}\frac{\partial T}{\partial r},   \\ 
                & \qquad 
      e^{-z/H} C = \left(f + \frac{2v}{r}\right)\left(f + \frac{\partial(rv)}{r\,\partial r}\right)
                 \equiv \hat{f}^2.   
  \end{split}
\label{eq4.5}
\end{equation}
Adding $\partial/\partial r$ of the second entry in (\ref{eq4.4}) to $\partial/\partial z$ 
of the first entry, we obtain the transverse circulation equation 
\begin{equation}                                           % Equation (4.6)
       \frac{\partial}{\partial r}
         \left( A \frac{\partial(r\psi)}{r\,\partial r} 
              + B \frac{\partial  \psi}{\partial z}\right)  
     + \frac{\partial}{\partial z}
         \left( B \frac{\partial(r\psi)}{r\,\partial r} 
              + C \frac{\partial  \psi}{\partial z}\right)      
     = \frac{g}{c_p T_0}\frac{\partial Q}{\partial r}.   
\label{eq4.6}
\end{equation}
Note that $ AC-B^2=(g/T_0)e^{(1-\kappa)z/H}\left(f+2v/r\right)P$, where 
\begin{equation}                                         % Equation (4.7)
      P = e^{z/H}\left[-\frac{\partial v}{\partial z}\frac{\partial\theta}{\partial r}    
        + \left(f + \frac{\partial(rv)}{r\,\partial r}\right)\frac{\partial\theta}{\partial z}\right]    
\label{eq4.7}     
\end{equation}
is the potential vorticity. The partial differential equation (\ref{eq4.6}) is 
elliptic if $(f+2v/r)P>0$. 

    In real tropical cyclones, the coefficients $A$, $B$, and $C$ can vary in complicated 
ways, making analytical solution of (\ref{eq4.6}) impossible. However, we can understand 
several aspects of the transverse circulation by solution of a simplified version of 
(\ref{eq4.6}). In this simplified version we assume $B=0$ and $A=e^{z/H} N^2$, where 
$N$ is a constant. Under these assumptions, (\ref{eq4.6}) simplifies to 
\begin{equation}                                           % Equation (4.8)
             N^2\frac{\partial}{\partial r}\left(\frac{\partial(r\psi)}{r\,\partial r}\right)  
     + \hat{f}^2\frac{\partial}{\partial z}\left(e^{z/H}\frac{\partial\psi}{\partial z}\right)      
     = \frac{ge^{-z/H}}{c_p T_0}\frac{\partial Q}{\partial r}.   
\label{eq4.8}
\end{equation}
It is important to note that the assumption $B=0$ precludes modeling the 
``stadium effect'' that was seen in Fig.~1. 
Assuming the diabatic heating $Q$, the streamfunction $\psi$, and the vertical 
velocity $w$ have the separable forms 
\begin{equation}                                         % Equation (4.9)
         Q(r,z) = \hat{Q}(r)    e^{ z/2H}\sin\left(\frac{\pi z}{z_T}\right), \quad   
      \psi(r,z) = \hat{\psi}(r) e^{-z/2H}\sin\left(\frac{\pi z}{z_T}\right), \quad   
         w(r,z) = \hat{w}(r)    e^{ z/2H}\sin\left(\frac{\pi z}{z_T}\right),            
\label{eq4.9}     
\end{equation}
the partial differential equation (\ref{eq4.8}) reduces to the ordinary 
differential equation 
\begin{equation}                                         % Equation (4.10)
        r^2 \frac{d^2\hat{\psi}}{dr^2}    
      + r   \frac{d  \hat{\psi}}{dr} 
      - \left(\frac{r^2}{\ell^2} + 1\right)\hat{\psi}
      = \frac{g r^2}{c_p T_0 N^2}\frac{d\hat{Q}}{dr},   
\label{eq4.10}     
\end{equation}
where 
\begin{equation}                                         % Equation (4.11)
     \hat{w} = \frac{d(r\hat{\psi})}{r\, dr}  \qquad \text{and} \qquad 
     \ell    = \frac{N}{\hat{f}}\left(\frac{\pi^2}{z_T^2} + \frac{1}{4H^2}\right)^{-1/2}.    
\label{eq4.11}     
\end{equation}
Since $\hat{f}$ is a function of radius, $\ell$ is also a function of radius and 
can be interpreted as the local Rossby length. 

\begin{figure}[tb]                               % Figure 8 (Bessel Functions)
\centering
\includegraphics[width=3.5in,clip=true]{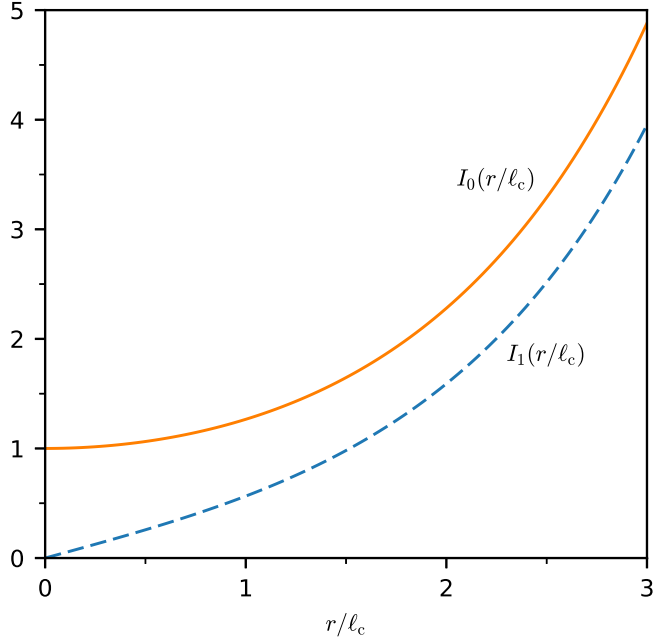}
\caption{The order-one modified Bessel function $I_1(r/\ell_c)$, from which 
the streamfunction $\hat{\psi}$ is constructed, and the order-zero modified Bessel 
function $I_0(r/\ell_c)$, from which the vertical motion $\hat{w}$ is constructed.}
\vspace*{1em}
\end{figure}

\begin{figure}[tb]                               % Figure 9 (Normalized w)
\centering
\includegraphics[width=4.5in,clip=true]{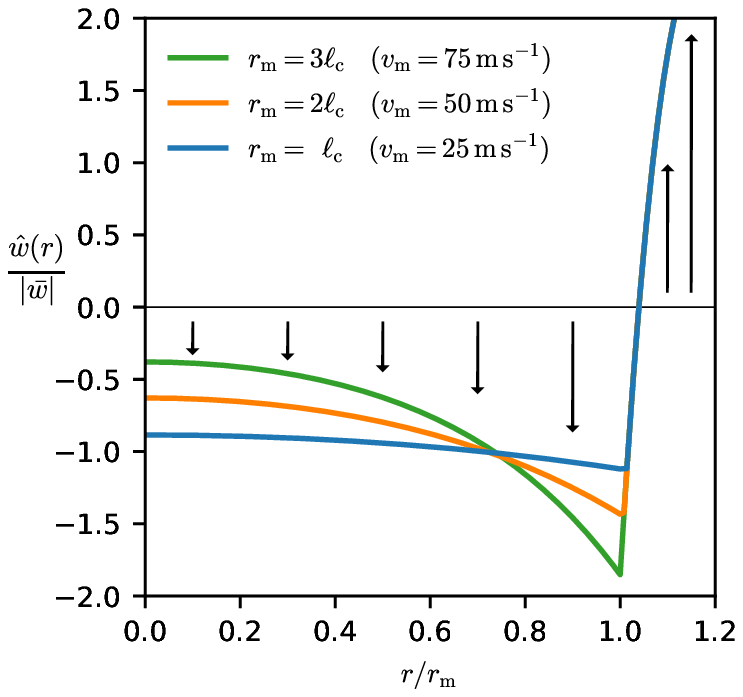}
\caption{Normalized vertical motion $\hat{w}(r)/|\bar{w}|$ as a function of radius, 
computed from (\ref{eq4.14}) for the eye region $0 \le r/r_{\rm m} < 1$ and 
simply shown schematically for the eyewall updraft region. The three colored 
curves have the same area-averaged subsidence over the eye-region. The 
$r_{\rm m}=3\ell_{\rm c}$ case (green curve) has the largest spatial variation 
of eye subsidence, with a subsidence rate at the outer edge of the eye nearly 
five times the value at the center.}
\vspace*{1em}
\end{figure}

     To solve (\ref{eq4.10}), we need to know the horizontal structure of the forcing 
$\hat{Q}(r)$ and the horizontal structure of the local Rossby length $\ell(r)$, and we 
need to enforce the boundary conditions $\hat{\psi}(0)=0$ and 
$r\hat{\psi}(r)\to 0$ as $r\to\infty$. However, to understand the distribution of 
subsidence in the eye, all we need to know is that, in the eye, equation (\ref{eq4.10}) 
becomes homogeneous and, to a good approximation, $\ell(r)$ takes on a constant 
``core value'' $\ell_c$. Thus, (\ref{eq4.10}) becomes 
\begin{equation}                                         % Equation (4.12)
        r^2 \frac{d^2\hat{\psi}}{dr^2}    
      + r   \frac{d  \hat{\psi}}{dr} 
      - \left(\frac{r^2}{\ell_c^2} + 1\right)\hat{\psi} = 0 
      \qquad  \text{for} \quad 0 \le r < r_{\rm m},   
\label{eq4.12}     
\end{equation}
where $r_{\rm m}$ is the radius of maximum tangential wind, which in the context 
of the present discussion can also be interpreted as the radius of the eye. The 
solution of (\ref{eq4.12}) is a linear combination of the order one modified Bessel 
functions $I_1(r/\ell_c)$ and $K_1(r/\ell_c)$. Because $K_1(r/\ell_c)$ is singular 
at $r=0$, only the $I_1(r/\ell_c)$ solution is accepted in the region 
$0\le r < r_{\rm m}$. Since $d[rI_1(r/\ell_c)]/dr=(r/\ell_c)I_0(r/\ell_c)$, the 
horizontal distribution of eye subsidence $\hat{w}(r)$ is proportional to the order 
zero modified Bessel function $I_0(r/\ell_c)$. Both $I_0(r/\ell_c)$ and 
$I_1(r/\ell_c)$ are plotted in Fig.~8. Since 
the transverse circulation problem is elliptic, the coefficient multiplying the 
$I_0(r/\ell_{\rm c})$ function depends on both the boundary conditions and 
the diabatic forcing over the entire domain. However, for the purposes of 
the present discussion, we are interested only in the relative spatial 
variation of the subsidence, not in its magnitude. Thus, we can simply 
normalize the eye solution through the 
use of $\bar{w}$, the area-averaged subsidence in the eye, defined by 
$\bar{w}=(2/r_{\rm m}^2)\int_0^{r_{\rm m}} \hat{w}(r)\, r\, dr$. Using this 
normalization, the eye solution can be expressed as 
\begin{equation}                                 % Equation (4.13)
   \frac{\hat{w}(r)}{|\bar{w}|} = -\frac{(r_{\rm m}/\ell_{\rm c})\, I_0(r/\ell_{\rm c})}
                                  {2I_1(r_{\rm m}/\ell_{\rm c})} 
			 \quad \text{for} \quad  0 \le r < r_{\rm m}.        
\label{eq4.13}
\end{equation}
The consistency of (\ref{eq4.13}) with the definition of $\bar{w}$ can 
easily be checked by taking the area integral of (\ref{eq4.13}) and using 
the integral relation $\int_0^x I_0(x')\, x' \, dx' = xI_1(x)$. 
Three useful special cases of (\ref{eq4.13}) are 
\begin{equation}                                 % Equation (4.14)
   \frac{\hat{w}(r)}{|\bar{w}|} 
    = \begin{dcases}
       -0.3794\, I_0(3r/r_{\rm m}) &\text{if } r_{\rm m}=3\ell_{\rm c}\,\,\,     (v_{\rm m}=75 ) \\
       -0.6287\, I_0(2r/r_{\rm m}) &\text{if } r_{\rm m}=2\ell_{\rm c}\,\,\,     (v_{\rm m}=50 ) \\
       -0.8847\, I_0( r/r_{\rm m}) &\text{if } r_{\rm m}= \ell_{\rm c}\,\,\,\,\, (v_{\rm m}=25).    
      \end{dcases}    
\label{eq4.14}
\end{equation}
Plots of (\ref{eq4.14}) are shown in Fig.~9. The three colored 
curves have the same area-averaged subsidence over the eye-region. The 
$r_{\rm m}=3\ell_{\rm c}$ case (green curve) has the largest spatial variation 
of eye subsidence, with a subsidence rate at the outer edge of the eye nearly 
five times the value at the center. Slow variation of the intensity of eyewall 
convection moves these eye-subsidence profiles up-and-down but does not alter 
their shape.

     The estimates of $v_{\rm m}$ given in (\ref{eq4.14}) are determined as follows. 
As a first approximation to the inner core radial structure of the azimuthal wind,  
we assume that $v(r)$ increases linearly with $r$ to the maximum value $v_{\rm m}$ at 
the radius $r_{\rm m}$. Then, $\hat{f}$ simplifies to $f+2v_{\rm m}/r_{\rm m}$ in the 
inner core, so that, according to (\ref{eq4.11}), the constant Rossby length in the 
core is given by   
\begin{equation}                                 % Equation (4.15)
   \ell_c = \left(\frac{f}{f + 2v_{\rm m}/r_{\rm m}}\right)\ell_0, 
                 \qquad \text{where} \qquad  
   \ell_0 = \frac{N}{f}\left(\frac{\pi^2}{z_T^2} + \frac{1}{4H^2}\right)^{-1/2} 
          = \frac{c}{f} \approx \frac{50\,\text{m}\,\text{s}^{-1}}{5\times 10^{-5}\,\text{s}^{-1}}
	    \approx 1000\text{ km}, 
\label{eq4.15}
\end{equation}
where $c$ is the gravity wave speed of the first internal mode. 
Using (\ref{eq4.15}), the ``dimensionless dynamical eye radius''  
$r_{\rm m}/\ell_c$ can be written as 
\begin{equation}                                 % Equation (4.16)
     \frac{r_{\rm m}}{\ell_c} 
   = \frac{\left(f + 2v_{\rm m}/r_{\rm m}\right)r_{\rm m}}{c}.  
\label{eq4.16}
\end{equation}
The orange curves in Fig.~10 are isolines of the core Rossby length $\ell_c$ in the 
$(r_{\rm m},v_{\rm m})$-plane. These isolines have been computed from 
(\ref{eq4.15}).  The blue curves in Fig.~10 are isolines of $r_{\rm m}/\ell_c$, 
i.e., the actual eye radius $r_{\rm m}$ measured in units of the core Rossby 
length $\ell_c$. The blue curves have been computed from (\ref{eq4.16}).
Note that the dimensionless dynamical eye radius $r_{\rm m}/\ell_c$ is 
essentially independent of the actual eye radius $r_{\rm m}$ and depends only 
on $v_{\rm m}$. In other words, to a first approximation, a set of tropical 
cyclones with the same values of $v_{\rm m}$ but widely varying actual eye 
sizes, all have essentially the same dynamical eye size and hence the same 
tendency to produce a warm-ring thermal structure. The weak dependence of 
$r_{\rm m}/\ell_c$ on $r_{\rm m}$ can be understood by noting that, 
as a vortex intensifies, $2v_{\rm m}/r_{\rm m}$ becomes much larger than $f$, 
so that the $f$ term in the numerator of (\ref{eq4.16}) can be neglected. Then 
the two $r_{\rm m}$ factors in (\ref{eq4.16}) cancel, yielding the approximation 
\begin{equation}                                 % Equation (4.17)
   \frac{r_{\rm m}}{\ell_c} \approx \frac{2v_{\rm m}}{c}
                            = \frac{v_{\rm m}}{25\, {\rm m}\, {\rm s}^{-1}}.    
\label{eq4.17}
\end{equation}
We conclude that, for practical purposes, the dimensionless dynamical eye radius  
is independent of the actual eye radius. The dimensionless dynamical eye radius  
depends only on $v_{\rm m}$.

\begin{figure}[tb]                               % Figure 10 (Eye Diameter)
\centering
\includegraphics[width=5.5in,clip=true]{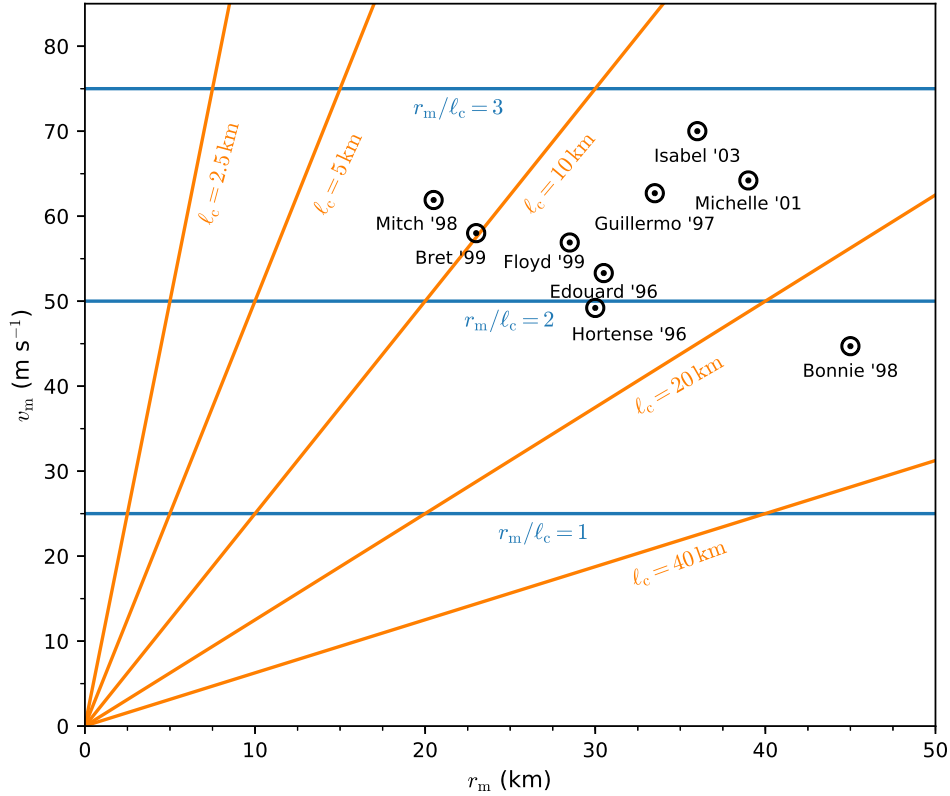}
\caption{The orange curves are isolines of the core Rossby length $\ell_c$, 
computed as a function of $r_{\rm m}$ and $v_{\rm m}$ from (\ref{eq4.15}). 
The blue curves are isolines of the dimensionless dynamical eye radius 
$r_{\rm m}/\ell_c$, calculated using (\ref{eq4.16}). The blue curves are 
accurately approximated by (\ref{eq4.17}). The black dots highlight examples 
of tropical cyclones with warm-ring thermal structures as observed by the 
NOAA WP-3D research aircraft.}
\vspace*{1em}
\end{figure}

     We can summarize this discussion with the following ``rules of thumb.'' 
For storms that have developed an eye but have not quite reached hurricane 
strength (blue curve in Fig.~9), the eye diameter is approximately two Rossby 
lengths and the subsidence is nearly uniform across the eye. For hurricanes that 
have reached an intensity of 50 m s$^{-1}$ (boundary of Saffir-Simpson 
Categories 2 and 3), the eye diameter is approximately four Rossby lengths 
and the subsidence at the edge of the eye is 2.3 times that at the center of 
the eye (red curve in Fig.~9).  For hurricanes that have reached an intensity 
of 75 m s$^{-1}$ (Category 5), the eye diameter is approximately six Rossby 
lengths and the subsidence at the edge of the eye is nearly five 
times that at the center of the eye (green curve in Fig.~9). Thus, 
simple balanced dynamics indicates that it is reasonable to expect that, 
no matter what their actual eye sizes, strong hurricanes should develop 
hub clouds, eye moats, and warm rings.  

    A survey of aircraft data taken in hurricanes with lower tropospheric warm 
rings reveals the nine cases indicated in Fig.~10. These hurricanes have eye 
diameters in the range $38 < 2r_{\rm m} < 90$ km, while their dimensionless 
dynamical eye diameters are generally in the range $4 < 2r_{\rm m}/\ell_c < 6$. 
It should be emphasized that the simple mathematical arguments given here 
neglect the baroclinic effects associated with eyewall slope and consider 
only the vertical structure of the first internal mode. Insightful discussions 
of the eye subsidence and thermal structure of numerically simulated tropical 
cyclones can be found in \citet{stern12} and \citet{stern13}.

\section{The tropical cyclone boundary layer}      %%%%%%%%%%  Section 5  %%%%%%%%%%

     We have seen in section 2 that a hollow PV tower will be produced 
if $\dot{\theta}$ is confined to an annular region. What dynamical process 
produces such a confinement? The answer appears to be that the nonlinear, 
frictional boundary layer dynamics of strong vortices dictate that intense 
boundary layer pumping, and hence large $\dot{\theta}$ in the overlying 
fluid, be confined to an annular region. Perhaps this is the dynamical 
process that Bernhard Haurwitz was seeking in the quote at the end of 
section 1. Although he did not discover the nonlinear dynamics discussed 
in this section, he did make an important contribution \citep{haurwitz35c,haurwitz36a}
to understanding the behavior of the frictional boundary 
layer under a specified, axisymmetric pressure field. This 
can be considered a classic problem in Ekman boundary layer theory,  
generalized in the sense that it involves highly curved flows with large 
Rossby numbers and with the possible formation of boundary layer 
shocks.\footnote{For an interesting discussion of how surface friction 
can lead to eyewall frontogenesis, see \citet{emanuel97}.} 
Here we simplify this problem by considering an axisymmetric slab 
boundary layer under a specified pressure field. We also consider a  
local, steady state model that is an approximation of the original model. 
The local model neglects the local time derivative terms, the horizontal 
diffusion terms, the vertical advection terms, and the radial advection of 
radial velocity; it also makes selective use of the gradient wind approximation.

    Both models consider axisymmetric, boundary layer motions of an 
incompressible fluid on an $f$-plane.  The frictional boundary layer 
is assumed to have constant density $\rho$ and constant depth $h$, 
with radial and azimuthal velocities $u(r,t)$ and $v(r,t)$ that 
are independent of height between the top of a thin surface layer 
and height $h$, and with vertical velocity $w(r,t)$ at height $h$.  In
the overlying layer, the radial velocity is assumed to be negligible and the
azimuthal velocity $v_{\rm gr}(r,t)$ is assumed to be in gradient balance 
and to be a specified function of radius and time. The boundary layer flow 
is driven by the same radial pressure gradient force that occurs in the 
overlying fluid. The governing system of 
differential equations for the boundary layer variables $u(r,t)$,
$v(r,t)$, and $w(r,t)$ then takes the form
\begin{equation}                                     % Equation (5.1)
      \frac{\partial u}{\partial t}
   + u\frac{\partial u}{\partial r} - \frac{w}{h}(1-\alpha)u
   - \left(f + \frac{v}{r}\right)v + \frac{1}{\rho}\frac{\partial p}{\partial r}
   = -\frac{\cD U}{h} u
   +  K\frac{\partial}{\partial r}\left(\frac{\partial(ru)}{r\partial r}\right),
\label{eq5.1}
\end{equation}
\begin{equation}                                     % Equation (5.2)
      \frac{\partial v}{\partial t}
   + u\frac{\partial v}{\partial r} - \frac{w}{h}(1-\alpha)\left(v-v_{\rm gr}\right)  
   + \left(f + \frac{v}{r}\right)u 
   = -\frac{\cD U}{h} v
   + K\frac{\partial}{\partial r}\left(\frac{\partial(rv)}{r\partial r}\right),
\label{eq5.2}
\end{equation}
\begin{equation}                                     % Equation (5.3)
      w = -h\frac{\partial(ru)}{r\partial r}  \quad {\rm and} \quad  
         \alpha = \begin{dcases} 
	                1  & {\rm if} \,\, w\ge 0  \\
			0  & {\rm if} \,\, w < 0,  
	          \end{dcases}
\label{eq5.3}
\end{equation}
where $f$ is the constant Coriolis parameter, $K$ the constant horizontal 
diffusivity, and $U = k\left(u^2 + v^2\right)^{1/2}$ the surface wind speed, 
with $k$ chosen to be 0.78. The 
boundary layer velocity field is driven by the specified pressure field 
$p(r,t)$ in (\ref{eq5.1}). However, it is more convenient to specify this 
forcing in terms of the associated gradient wind field $v_{\rm gr}(r,t)$, 
which is related to $p(r,t)$ via the gradient wind formula  
\begin{equation}                                  % Equation (5.4)
     \left(f + \frac{v_{\rm gr}}{r}\right)v_{\rm gr}
     = \frac{1}{\rho}\frac{\partial p}{\partial r}.                    
\label{eq5.4}
\end{equation}

     In the absence of the horizontal diffusion terms, the slab boundary 
layer equations constitute a hyperbolic system that can be written 
in characteristic form \citep{slocum14}. A knowledge of the characteristic 
form is useful in understanding the formation of shocks. Before presenting 
numerical solutions of the system (\ref{eq5.1})--(\ref{eq5.3}), 
we present a local model which proves to give accurate results at all radii 
except near the radius of shock formation.  

     The local, steady state approximation is derived by first neglecting 
the $(\partial/\partial t)$-terms, the horizontal diffusion terms, and the 
$w$-terms in (\ref{eq5.1}) and (\ref{eq5.2}). The resulting equations can 
then be written in the form\footnote{Interesting discussions of steady state 
models similar to (\ref{eq5.5}) are given by \citet{smithvogl08} 
and \citet{kepert10a,kepert10b}.}      
\begin{equation}                                    % Equation (5.5)
  \begin{split}
     &u\frac{\partial u}{\partial r} 
                     -(f+\zeta)v + \frac{\cD U}{h}u 
	    = -\frac{\partial}{\partial r}\left(\frac{p}{\rho}+\tfrac{1}{2}v^2\right), \\
     & \qquad\qquad   (f+\zeta)u + \frac{\cD U}{h}v = 0,  
  \end{split}
\label{eq5.5}
\end{equation}
where the boundary layer relative vorticity is defined by 
$\zeta = \partial(rv)/r\,\partial r$. 
Next, approximate $f+\zeta$ by $f+\zeta_{\rm gr}$, and approximate 
$(p/\rho)+\frac{1}{2}v^2$ by $(p/\rho)+\frac{1}{2}v_{\rm gr}^2$, thereby obtaining 
\begin{equation}                                    % Equation (5.6)
  \begin{split}
       u\frac{\partial u}{\partial r} - (f+\zeta_{\rm gr})v + \frac{\cD U}{h}u 
                                                &= -(f+\zeta_{\rm gr})v_{\rm gr},  \\
	 (f+\zeta_{\rm gr})u + \frac{\cD U}{h}v &= 0,  
  \end{split}
\label{eq5.6}
\end{equation}
where the relative vorticity of the gradient wind is defined by 
$\zeta_{\rm gr} = \partial(rv_{\rm gr})/r\,\partial r$,  
and where we have made use of (\ref{eq5.4}).  The final step in the derivation 
of the local model is to neglect the $u(\partial u/\partial r)$ term in the top 
line of (\ref{eq5.6}). This removes the nonlocal nature of (\ref{eq5.6}) and results 
in the local model equations 
\begin{equation}                                    % Equation (5.7)
  \begin{split}
     -(f+\zeta_{\rm gr})v + \left(\frac{k\cD (u^2+v^2)^{1/2}}{h}\right)u 
	                               &= -(f+\zeta_{\rm gr})v_{\rm gr},  \\
      (f+\zeta_{\rm gr})u + \left(\frac{k\cD (u^2+v^2)^{1/2}}{h}\right)v 
                                       &= 0.   
  \end{split}
\label{eq5.7}
\end{equation}
Given a radial profile $v_{\rm gr}(r)$ and its associated vorticity 
$\zeta_{\rm gr}(r)$, we can solve the nonlinear equations (\ref{eq5.7}) 
for $u$ and $v$ at each radial point. 
\begin{figure}[tb]                           % Figure 11 (Gradient Wind Profiles)
\centering
\includegraphics[width=4.5in,clip=true]{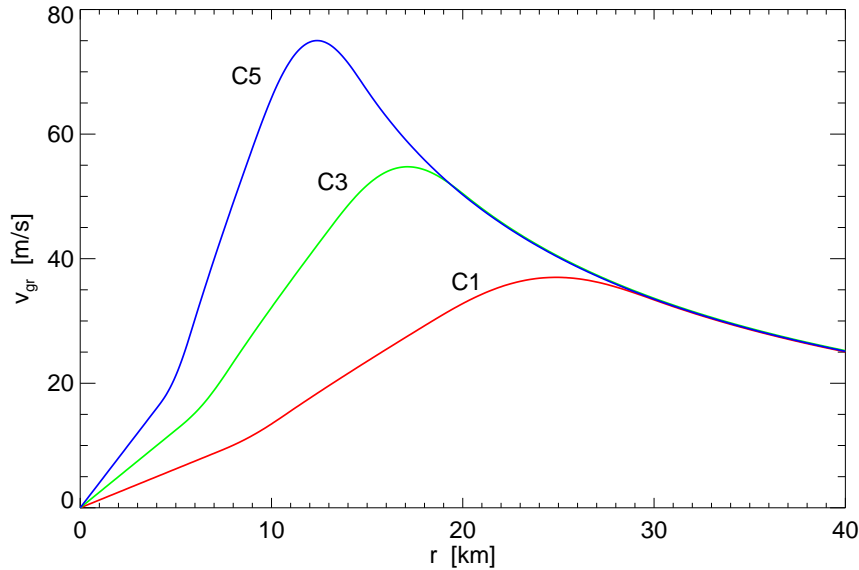}
\caption{Radial profiles of $v_{\rm gr}$ for the category 1, 3, and 5 cases. 
From \citet{williams13}.}
\vspace*{1em}
\end{figure}

\begin{figure}[tb]                           % Figure 12 (Slab Boundary Layer Results)
\centering
\includegraphics[width=5.5in,clip=true]{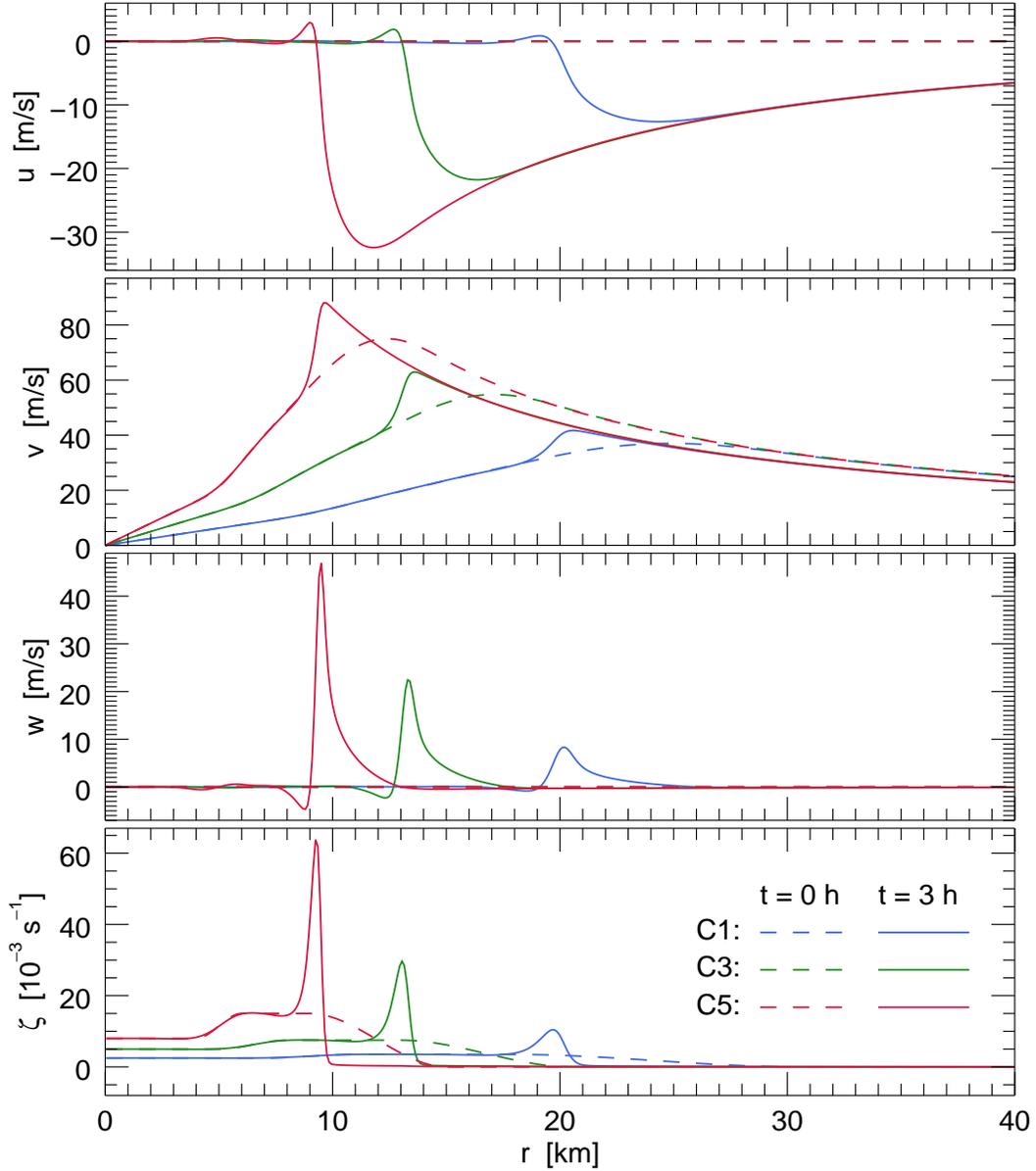}
\caption{Steady-state (i.e., $t=3$ hours) slab boundary layer results for the 
C1, C3, and C5 forcing cases. From \citet{williams13}.}
\vspace*{1em}
\end{figure}

     Figure 11 shows $v_{\rm gr}(r)$ profiles for the three vortices studied 
by \citet{williams13}. The maximum values of the gradient 
wind $v_{\rm gr}(r)$ for the Category 1, 3, and 5 cases are 37.5, 55, and 
75 m~s$^{-1}$, respectively. Numerical integration\footnote{The formula used 
to describe the dependence of $\cD$ on $U$ is given in \citet{williams13}.} 
of (\ref{eq5.1})--(\ref{eq5.3}) 
shows that in all three cases, the vortices quickly evolve into 
a nearly steady state, with small changes after 2 or 3 hours.  
Figure~12 shows $u$, $v$, $w$, and $\zeta$ for $t=0$ and $t=3$ hours for
each category vortex.  An interesting feature of the radial profiles of
$w(r,t)$ is the very sharp gradient on the inner side and the relatively
slower decrease on the outer side of the maximum. Through the mass continuity
equation (\ref{eq5.3}), this behaviour of $w(r,t)$ is related to the shock-like 
structure\footnote{Here we are using the term ``shock'' rather than the term 
``front'', because the near discontinuity in $u$ arises from 
$u(\partial u/\partial r)$, i.e., the advection of the divergent wind component 
by the divergent wind component. This term is neglected in the accepted 
semi-geostrophic theory of line-symmetric atmospheric fronts.} 
of $u(r,t)$.  To a certain extent, the radial structure of $w(r,t)$
in the slab model agrees with the observed radial structure in the famous 
Hurricane Hugo case, which is shown in Fig.~13. 
Another interesting feature of the Category 3 and
5 cases is the rather large boundary layer pumping that the slab model
produces, when compared with the observations of Fig.~13. Some of this
discrepancy is probably due to the simplicity of the slab model and the chosen
value of $K$, but some may be explained by the low elevation of the flight level
for the red curve in the lower panel of Fig.~13, i.e., there may have been 
radially convergent flow above flight level so that a larger $w$ might have 
been measured if the aircraft had flown several hundred meters higher. 
In actual hurricanes, this shock effect makes the inner core boundary layer a
dangerous place for research aircraft. The location of shock formation also
plays a crucial role in determining the eyewall radius, and hence where the
diabatic heating will occur relative to the region of high inertial stability. 

     Although not shown here, the solutions of the local model (\ref{eq5.7}) 
are close to the solutions of the non-local model (\ref{eq5.1})--(\ref{eq5.3}), 
except near the region of the shock. In this region, the local model is 
inaccurate, primarily because of the neglect of the $u(\partial u/\partial r)$ 
term. Thus, it is the nonlinear dynamics of the boundary layer that determines 
the location of the eyewall and hence the diameter of the eye.

\begin{figure}[tb]                               % Figure 13 (Hurricane Hugo)
\centering
\includegraphics[width=5.5in,clip=true]{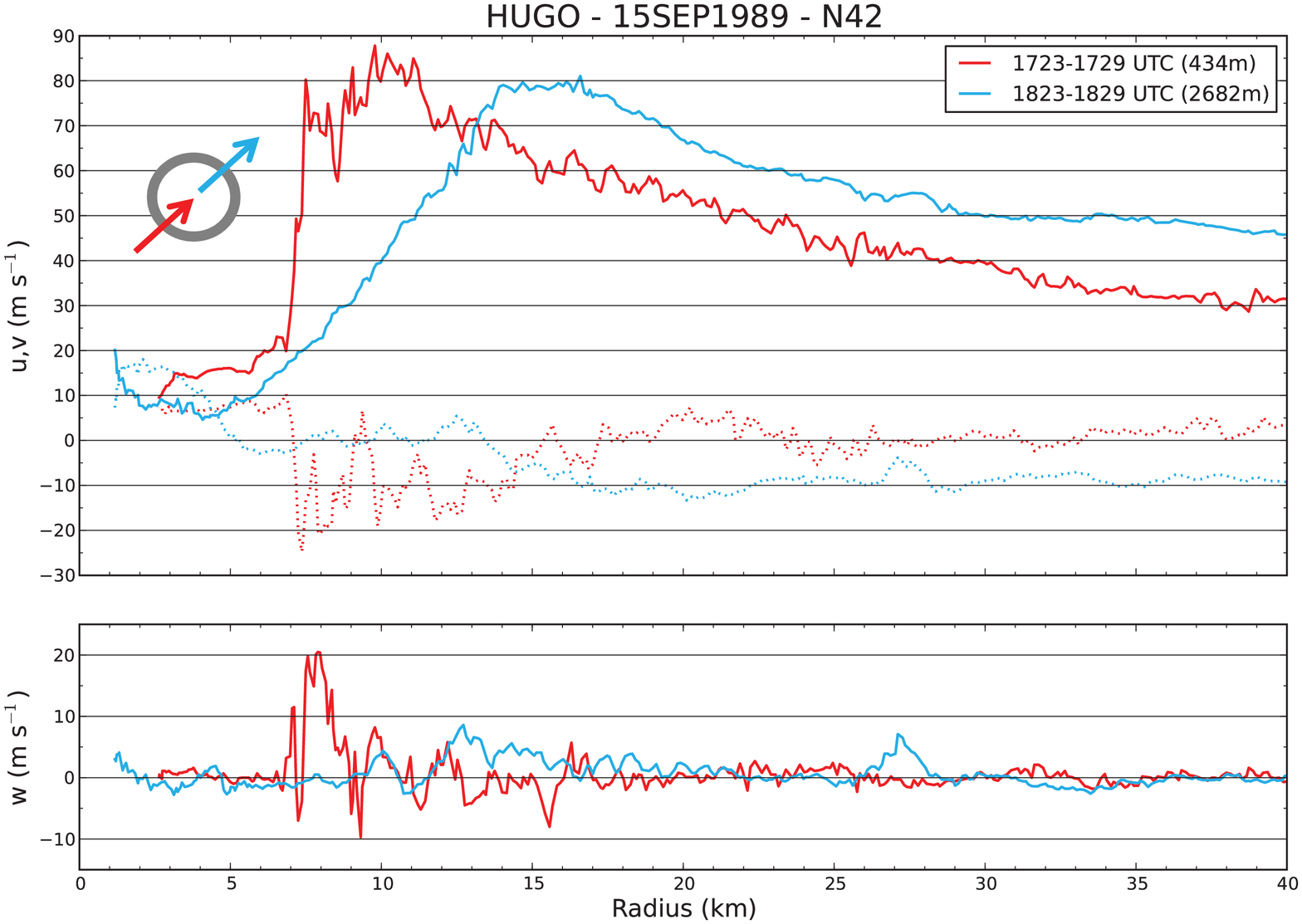}
\caption{The top panel shows NOAA WP-3D aircraft data from an inbound leg 
in the southwest quadrant (red, 434 m average height) and an outbound leg 
in the northeast quadrant (blue, 2682 m average height) for Hurricane Hugo 
on 15 September 1989. The solid curves show the tangential wind component, 
while the dotted curves show the radial wind component. The bottom panel 
shows the vertical velocity. The shock-like feature in the boundary layer 
(434 m height) occurs at a radius of 7--8 km. From \citet{williams13}.}
\vspace*{1em}
\end{figure}

\section{The formation of PV anomalies in the ITCZ}   %%%%%  Section 6  %%%%%

     In this section, we discuss the strong analogy between the PV dynamics 
of the ITCZ and the PV dynamics of a tropical cyclone. As in a tropical 
cyclone, there are transient inertia-gravity waves in the ITCZ, 
but a useful idealization is that the ITCZ is essentially a balanced, 
PV phenomenon, with the balanced wind and mass fields invertible from 
the PV field. Again, with frictional effects confined primarily to a 
shallow boundary layer, the evolution of PV above the boundary layer 
is determined by advection and diabatic effects. Considering zonally symmetric, 
inviscid flow on the sphere, and using latitude $\phi$, height $z$, and time $t$ 
as the independent variables, the PV equation takes the form  
\begin{equation}                                      % Equation (6.1)
       \frac{\partial P}{\partial t}
    + v\frac{\partial P}{a\partial\phi}
    + w\frac{\partial P}{\partial z} 
    = \frac{1}{\rho}\left[
      \frac{\partial u}{\partial z}\frac{\partial\dot{\theta}}{a\partial\phi}
    + \left(2\Omega\sin\phi - \frac{\partial(u\cos\phi)}{a\cos\phi\,\partial\phi}\right)
              \frac{\partial\dot{\theta}}{\partial z}\right],  	  	  
\label{eq6.1}
\end{equation} 
where $u$ is the zonal velocity, $v$ the meridional velocity, $w$ the vertical 
velocity, $\rho$ the density, $\dot{\theta}$ the diabatic heating, and where 
the potential vorticity is given by 
\begin{equation}                                      % Equation (6.2)
     P = \frac{1}{\rho}\left[
         \frac{\partial u}{\partial z}\frac{\partial\theta}{a\partial\phi}
       + \left(2\Omega\sin\phi - \frac{\partial(u\cos\phi)}{a\cos\phi\,\partial\phi}\right)
                 \frac{\partial\theta}{\partial z}\right]
       = \frac{2\Omega\sin\Phi}{\rho }\frac{\partial(\sin\Phi,\theta)}{\partial(\sin\phi,z)},   	  	  
\label{eq6.2}
\end{equation} 
with the potential latitude $\Phi$ defined in terms of the absolute angular 
momentum by $\Omega a\cos^2\Phi=u\cos\phi+\Omega a\cos^2\phi$. We will now collapse 
the partial differential equation (\ref{eq6.1}) into two simple ordinary 
differential equations, which can be solved analytically. To accomplish 
this simplification, first transform from the original independent variables 
$(\phi,z,t)$ to the new independent variables $(\Phi,\theta,\tau)$, where $\tau=t$, 
but $\partial/\partial\tau$ means that $\Phi$ and $\theta$ are fixed, 
while $\partial/\partial t$ means that $\phi$ and $z$ are fixed. Then, 
the $(\phi,z,t)$-form of the material derivative is related the  
$(\Phi,\theta,\tau)$-form by 
\begin{equation}                                      % Equation (6.3)
                  \frac{\partial}{\partial t}   
               + v\frac{\partial}{a\partial\phi} 
               + w\frac{\partial}{\partial z} 
               =  \frac{\partial}{\partial\tau} 
      + \dot{\Phi}\frac{\partial}{\partial\Phi}
    + \dot{\theta}\frac{\partial}{\partial\theta},  	  	  
\label{eq6.3}
\end{equation} 
where $\dot{\Phi}$ is the rate that fluid particles are crossing $\Phi$-surfaces. 
For inviscid zonally symmetric flow, the absolute angular momentum is materially 
conserved, so that $\dot{\Phi}=0$, which simplifies the right-hand side of 
(\ref{eq6.3}) and hence the left-hand side of (\ref{eq6.1}). The right-hand 
side of (\ref{eq6.1}) can also be simplified by writing it as 
\begin{equation}                                      % Equation (6.4)
  \begin{split}
    & \frac{1}{\rho}\left[
      \frac{\partial u}{\partial z}\frac{\partial\dot{\theta}}{a\partial\phi}
    + \left(2\Omega\sin\phi - \frac{\partial(u\cos\phi)}{a\cos\phi\,\partial\phi}\right)
              \frac{\partial\dot{\theta}}{\partial z}\right]   \\   
    & \qquad \qquad \qquad 
     = \frac{2\Omega\sin\Phi}{\rho}
                        \frac{\partial(\sin\Phi,\dot{\theta})}{\partial(\sin\phi,z)}   
     = \frac{2\Omega\sin\Phi}{\rho}
                        \frac{\partial(\sin\Phi,\theta)}{\partial(\sin\phi,z)} 
                        \frac{\partial(\sin\Phi,\dot{\theta})}{\partial(\sin\Phi,\theta)}
     = P\frac{\partial\dot{\theta}}{\partial\theta},  	  	  
  \end{split}
\label{eq6.4}
\end{equation} 
where the second equality makes use of the Jacobian chain rule and the 
third equality makes use of (\ref{eq6.2}). Then, using (\ref{eq6.3}) and 
(\ref{eq6.4}) in (\ref{eq6.1}), the PV equation becomes 
\begin{equation}                                      % Equation (6.5)
                  \frac{\partial P}{\partial\tau} 
    + \dot{\theta}\frac{\partial P}{\partial\theta} 
    = P\frac{\partial\dot{\theta}}{\partial\theta}.   	  	  
\label{eq6.5}
\end{equation} 
The PV dynamics (\ref{eq6.5}), which is equivalent to (\ref{eq6.1}), can 
now be written in the characteristic form 
\begin{equation}                                      % Equation (6.6)
      \frac{d\ln P}{d\tau} = \frac{\partial\dot{\theta}}{\partial\theta}
                \quad {\rm on} \quad 
      \frac{d\theta}{d\tau} = \dot{\theta},   	  	  
\label{eq6.6}
\end{equation} 
where $(d/d\tau)=(\partial/\partial\tau)+\dot{\theta}(\partial/\partial\theta)$
is the derivative along the characteristic. As in the tropical cyclone case 
(section 2) the pair of ordinary differential equations in (\ref{eq5.6}) can be 
solved sequentially, i.e., the second can be solved for the shapes of the 
characteristics and then the first can be solved for the variation of PV along 
the characteristics. Since (\ref{eq6.5}) is identical to (\ref{eq2.5}), and 
since (\ref{eq6.6}) is identical to (\ref{eq2.6}), the PV dynamics 
along a potential radius surface in a tropical cyclone is identical 
to the PV dynamics along a potential latitude surface in the ITCZ. However, two 
important differences are the inclusion of variable Coriolis parameter in the 
ITCZ case and the rates at which the PV evolves. Since the values of 
$\dot{\theta}$ in a tropical cyclone eyewall are so much larger than in the ITCZ, 
the $\tau_c$-clock for an $R$-surface in a tropical cyclone eyewall generally runs 
much faster than the $\tau_c$-clock for a $\Phi$-surface in the ITCZ.

     For the example considered here, we assume that $\dot{\theta}$ is 
independent of $\tau$ and has the separable form 
\begin{equation}                                      % Equation (6.7)
   \dot{\theta}(\Phi,\theta) = \dot{\Theta}(\Phi) 
		\sin^2\left(\frac{\pi(\theta-\theta_B)}{\theta_T-\theta_B}\right),     	  	  
\label{eq6.7}
\end{equation} 
where $\dot{\Theta}(\Phi)$ is a specified function, and where the bottom and 
top isentropes are again chosen as $\theta_B=300\,$K and $\theta_T=360\,$K. 
This $\sin^2$ profile produces PV anomalies that are entirely internal since 
$\partial\dot{\theta}/\partial\theta$ vanishes at $\theta=\theta_B,\theta_T$. 
This is probably more appropriate when there is a significant fraction of 
stratiform precipitation that produces evaporative cooling near the surface. 

    Using (\ref{eq6.7}) in the second equation of (\ref{eq6.6}), we obtain 
\begin{equation}                                      % Equation (6.8)
    \frac{\left(\frac{\pi}{\theta_T-\theta_B}\right)d\theta}
         {\sin^2\left(\frac{\pi(\theta-\theta_B)}{\theta_T-\theta_B}\right)} 
    = d\tau_c,       \qquad \text{where} \qquad   	  	  
        \tau_c = \frac{\pi\dot{\Theta}(\Phi)\tau}{\theta_T-\theta_B}.  
\label{eq6.8}
\end{equation} 
As in the hurricane case, this dimensionless convective clock runs at different 
rates on different $\Phi$-surfaces. In 
the center of the ITCZ, where $\dot{\Theta}(\Phi)$ is large, the $\tau_c$-clock 
advances quickly, while near the edge of the ITCZ, the $\tau_c$-clock 
advances slowly, or not at all. Integration of (\ref{eq6.8}) yields the 
characteristic equation  
\begin{equation}                                      % Equation (6.9)
   \theta(\vartheta,\tau_{\rm c}) = \theta_B + \frac{(\theta_T-\theta_B)}{\pi}
      \cot^{-1}\left[  
      \cot\left(\frac{\pi(\vartheta-\theta_B)}{\theta_T-\theta_B}\right) - \pi\tau_c\right], 
\label{eq6.9}
\end{equation} 
where $\vartheta$ is the label of the characteristic, i.e., the initial 
potential temperature of the characteristic. To confirm that (\ref{eq6.9}) 
satisfies the initial condition, note that the $\pi\tau_c$ factor becomes 
zero when $\tau_c=0$ and that the inverse cotangent and cotangent operations then 
cancel, so that (\ref{eq6.9}) reduces to $\theta=\vartheta$ when $\tau_c=0$. 
The red curves in Fig.~14 are the characteristics $\theta(\vartheta,\tau_{\rm c})$ 
given by (\ref{eq6.9}). These characteristics bend upward most rapidly at 
$\theta=330\,$ K, where the value of $\dot{\theta}$ is largest. As in section 2, 
it is useful to rearrange (\ref{eq6.9}) into the form 
\begin{equation}                                      % Equation (6.10)
    \vartheta(\theta,\tau_c) = \theta_B 
        + \frac{\theta_T-\theta_B}{\pi} \cot^{-1}\left[  
	  \cot\left(\frac{\pi(\theta-\theta_B)}{\theta_T-\theta_B}\right) + \pi\tau_c\right],     	  
\label{eq6.10}
\end{equation} 
which can be regarded as giving the initial potential temperature of the characteristic 
that goes through the point $(\theta,\tau_c)$. 

     We now turn our attention to the solution of the first ordinary differential 
equation in (\ref{eq6.6}). With the previous assumption (\ref{eq6.7}) that 
$\dot{\theta}$ depends only on $(\Phi,\theta)$, and now assuming that the initial 
potential vorticity is $P_0\sin\Phi$, where $P_0$ is a constant, the solution of 
the first equation in (\ref{eq6.6}) is 
\begin{equation}                                      % Equation (6.11)
   P(\Phi,\theta,\tau_c) = P_0 \sin\Phi 
                     \left(\frac{\dot{\theta}(\Phi,\theta)}
		                {\dot{\theta}(\Phi,\vartheta(\theta,\tau_c))}\right).    	  	  
\label{eq6.11}
\end{equation} 
Using (\ref{eq6.7}) and (\ref{eq6.10}) in (\ref{eq6.11}), we obtain the final solution 
\begin{equation}                                      % Equation (6.12)
   \frac{P(\Phi,\theta,\tau_c)}{P_0} = \sin\Phi \left( 
          \frac{\sin^2\left(\frac{\pi(\theta-\theta_B)}{\theta_T-\theta_B}\right)}
	       {\sin^2\left\{\cot^{-1}\left[  
	           \cot\left(\frac{\pi(\theta-\theta_B)}{\theta_T-\theta_B}\right) 
		                + \pi\tau_c(\Phi) \right]\right\}}\right).    	  	  
\label{eq6.12}
\end{equation} 
Although the right-hand side of (\ref{eq6.12}) is indeterminant at the boundaries,  
use of L'Hospital's rule yields $P(\theta,\Phi,\tau_c)=P_0 \sin\Phi$ at 
$\theta=\theta_B,\theta_T$.  
\begin{figure}[tb]                                  % Figure 14
\centering
\includegraphics[width=5.5in,clip=true]{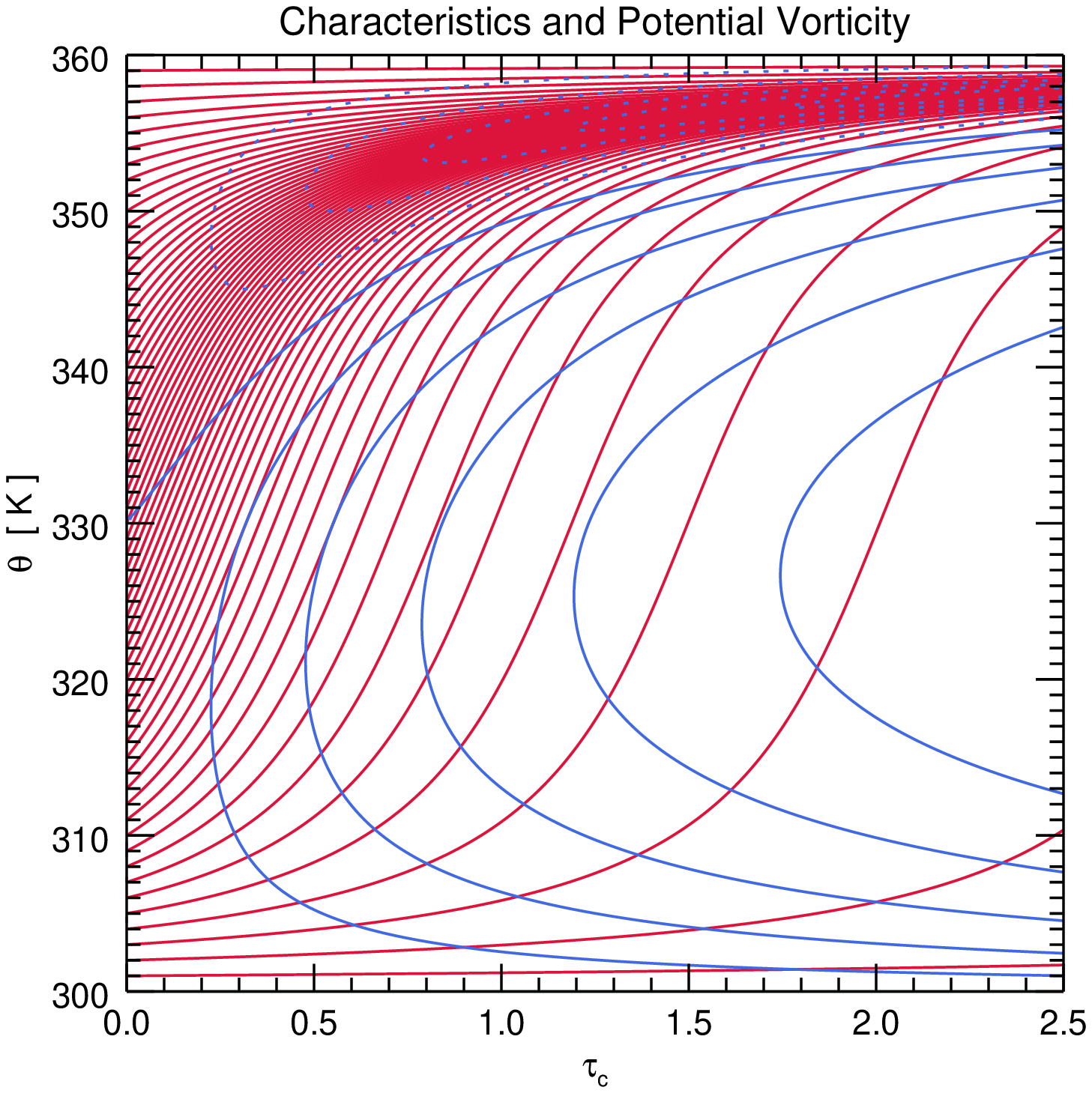}
\caption{The red curves are the characteristics $\theta(\vartheta,\tau_{\rm c})$ 
determined from (\ref{eq6.9}), with $\vartheta$ denoting the characteristic label 
(i.e., the initial value of the potential temperature) and $\tau_{\rm c}$ denoting 
the dimensionless convective clock defined in the second entry of (\ref{eq6.8}). 
The blue curves are isolines of $P(\theta,\tau_{\rm c})/P_0\sin\Phi$, as determined from 
(\ref{eq6.12}). The blue isoline starting at 330\,K corresponds to 
$P(\theta,\tau_{\rm c})/P_0\sin\Phi=1$, with the other solid isolines corresponding to 
$P(\theta,\tau_{\rm c})/P_0\sin\Phi=2,4,8,16,32$ and the dashed lines corresponding to 
$P(\theta,\tau_{\rm c})/P_0\sin\Phi=1/2,1/4,1/8,1/16$.}
\vspace*{1em}
\end{figure}

    Isolines of the dimensionless potential vorticity $P(\theta,\Phi,\tau_c)/P_0\sin\Phi$ 
are shown by the blue curves in Fig.~14.  The largest values of PV occur at 
$\tau_{\rm c}=2.5$ on midtropospheric isentropes. In the region $300 \le \theta\le 330\,$K,
both the $\dot{\theta}(\partial P/\partial\theta)$ and the 
$P(\partial\dot{\theta}/\partial\theta)$ terms in (\ref{eq6.5}) contribute 
to positive $(\partial P/\partial\tau)$. In the region $330 \le \theta\le 360\,$K,
the $P(\partial\dot{\theta}/\partial\theta)$-term makes a negative contribution 
to $(\partial P/\partial\tau)$, but $(\partial P/\partial\tau)$ remains 
positive because of the $\dot{\theta}(\partial P/\partial\theta)$-term. The 
net effect is that a tower of high PV grows into the upper troposphere. 

\begin{figure}[tb]                                  % Figure 15
\centering
\includegraphics[width=5.5in,clip=true]{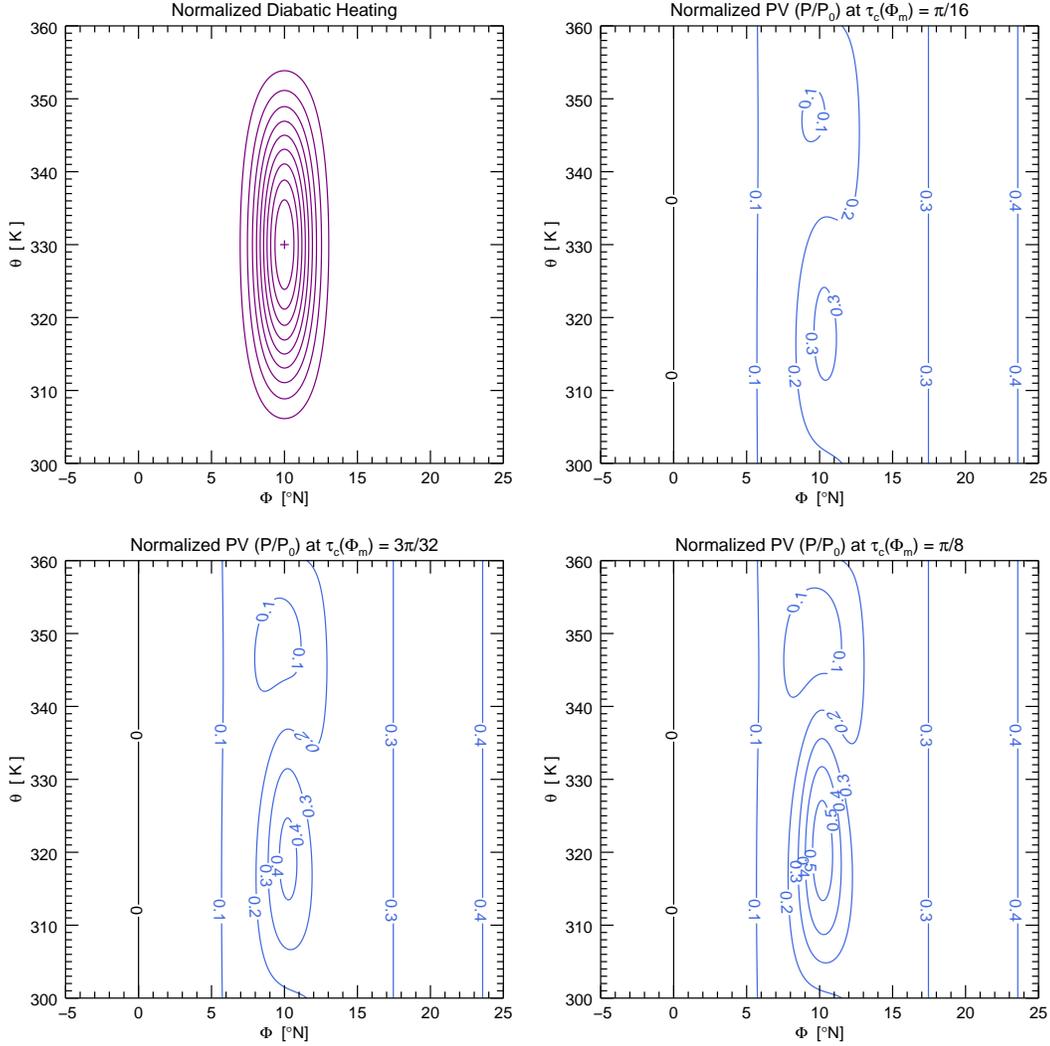}
\caption{The upper left panel shows isolines of 
$\dot{\theta}(\Phi,\theta)/\dot{\Theta}_m$. The remaining three panels show 
isolines of $P(\Phi,\theta,\tau_{\rm c})/P_0$, as determined from 
(\ref{eq6.12}) and (\ref{eq6.13}), for $\Phi_m=10^\circ$, $\Phi_w=2^\circ$, 
and for the three times corresponding to $\tau_c(\Phi_m)=\pi/16,3\pi/32,\pi/8$.}
\vspace*{1em}
\end{figure}

     To get a better appreciation of the formation of a PV anomaly in the 
ITCZ, consider the example $\dot{\Theta}(\Phi)=\dot{\Theta}_m \exp[-(\Phi-\Phi_m)^2/\Phi_w^2]$, 
where $\Phi_m$ is the potential latitude of the maximum diabatic heating (center of the ITCZ), 
$\Phi_w$ is the width of the diabatic heating, and $\dot{\Theta}_m$ is the maximum value of 
$\dot{\Theta}(\Phi)$. Using this $\dot{\Theta}(\Phi)$ in the second entry of 
(\ref{eq6.8}), we obtain  
\begin{equation}                                      % Equation (6.13)
     \tau_c(\Phi) = \tau_c(\Phi_m)  
                 \exp\left(-\frac{(\Phi-\Phi_m)^2}{\Phi_w^2}\right), 
              \qquad \text{where} \qquad 
     \tau_c(\Phi_m) = \frac{\pi\dot{\Theta}_m \tau}{\theta_T-\theta_B}.            	  	  
\label{eq6.13}
\end{equation} 
Using (\ref{eq6.13}) in (\ref{eq6.12}), we can now produce $(\Phi,\theta)$-cross-sections 
of PV at different times. Figure 15 shows isolines of PV in $(\Phi,\theta)$-space for the 
choice $\Phi_m=10^\circ$, $\Phi_w=2^\circ$ and for the three times 
$\tau_c(\Phi_m)=\pi/16,3\pi/32,\pi/8$. Note that a mini-tower of PV is 
produced.\footnote{For additional discussion, see \citet{schubert91}.}

    Since zonally symmetric PV anomalies develop within a background state that 
has potential vorticity increasing to the north, reversed poleward gradients 
of potential vorticity are produced. For typical diabatic heating rates, 
significant PV gradient reversals can occur within a couple of days. This 
sets the stage for combined barotropic-baroclinic instability and the growth 
of easterly waves.  The barotropic aspects of this instability process are 
discussed by \citet{ferreira97}, who used a global 
shallow water model. More detailed simulations have been made by
\citet{magnusdottir08} and \citet{wang05,wang06}, 
using a multi-level global spectral model. In a companion observational study   
\citet{bain11} have developed objective methods for detecting the 
ITCZ using instantaneous satellite data and have produced an ITCZ climatology 
for the east Pacific.  

     It should be noted that this PV interpretation of the ITCZ and Hadley 
circulation is quite different than the interpretations of 
\citet{schneider+lindzen77}, \citet{schneider77}, 
\citet{held80}, and \citet{lindzen88}, 
all of whom seek steady state solutions for the forced, zonally symmetric flow. 
In the present analysis, the diabatic heating $\dot{\theta}$ results in a 
continuously evolving PV field, so a steady state is never reached. However, 
the evolving flow field does become unstable to combined barotropic-baroclinic 
instability.

\section{Concluding remarks}                       %%%%%  Section 8  %%%%%

     The concepts presented here, involving the formation of PV towers in both 
tropical cyclones and the ITCZ, are quite different from those presented a half 
century ago by \citet{charney64} and \citet{charney71}.
The older view was that the basic dynamics of 
both tropical cyclones and the ITCZ could be understood through linear 
stability analyses involving balanced models with parameterized convection. 
For unstable disturbances, the primary circulation, the secondary circulation, 
and the diabatic heating all increase exponentially with time. The modern view 
is that a better understanding is obtained through analysis of the nonlinear 
aspects of the PV evolution. In this view the primary circulation can increase 
exponentially in time while the diabatic heating is fixed. The modern view also 
recognizes that the nonlinear dynamics of the boundary layer produces the sharp, 
intense boundary layer pumping that dictates the position of the eyewall and 
the size of the eye. A major challenge for future research is to improve intensity 
forecasting by improving numerical simulation of the tropical cyclone's mesoscale 
power plant. Just how much improvement is possible remains unknown, since there 
are no doubt some basic limitations to the predictability of tropical cyclone 
intensity. 

     For those interested in further information about the life of 
Bernhard Haurwitz, there are several excellent sources. The first is his 
own series of four short articles entitled ``Meteorology 
in the 20th Century---A Participant's View,'' published in the Bulletin of 
the American Meteorological Society \citeyearpar{haurwitz85}. The second is the 
National Academy of Sciences Biographical Memoir written by his longtime 
colleague Julius London \citeyearpar{london96}.  The third is an extensive
transcribed interview entitled ``Conversations with Bernhard Haurwitz,'' 
conducted by master interviewer George Platzman \citeyearpar{platzman85}. 
Finally, there is an excellent historical article, also by Platzman
\citeyearpar{platzman96},
on the diurnal tide in surface pressure, entitled ``The S-1 Chronicle:
A Tribute to Bernhard Haurwitz.''

\section*{Acknowledgements}

I would like to thank Rick Taft and Chris Slocum for their help in 
preparing this Bernhard Haurwitz Memorial Lecture. I would also like 
to thank the AMS for their continuing support of this lecture series 
since its inception in 1989. The author's research is supported by 
the National Science Foundation under Grants AGS-1546610 and AGS-1601623.  

\bibliographystyle{ametsoc2014}
\bibliography{references}

\end{document}